\documentclass{aa}  

\usepackage{graphicx}
\usepackage{natbib}
\usepackage{longtable}

\usepackage{multicol, blindtext}
\usepackage{txfonts}
\begin{document}

   \title{Evidence of internal rotation and a helical magnetic field in the jet of the quasar NRAO\,150} %


   \author{    Sol N. Molina\inst{1}
             \and 
                   Iv\'an Agudo\inst{1,2,3}
              \and 
                    Jos\'e L. G\'omez\inst{1}
               \and 
                     Thomas P. Krichbaum\inst{4}
               \and
                     Iv\'an Mart\'i-Vidal\inst{5}
               \and 
                     Alan L. Roy\inst{4}}
 
              \institute{Instituto de Astrof\'isica de Andaluc\'ia, CSIC, Apartado 3004, 18080 
              Granada, Spain; \email{smolina@iaa.es, jlgomez@iaa.es}
              \and
               Institute for Astrophysical Research, Boston University, 725 Commonwealth Avenue, 
               Boston, MA 02215-1401, USA.
               \and
               Current Address: Joint Institute for VLBI in Europe, Postbus 2, NL-7990 AA, 
               Dwingeloo, the Netherlands; \email{agudo@jive.nl}
               \and
                Max-Planck-Institut f\"ur Radioastronomie, Auf dem H\"ugel, 69, D-53121, 
                Bonn, Germany
                \and
                Chalmers University of Technology, Onsala Space Observatory, SE-43992 
                Onsala, Sweden} 

 
  \abstract{
The source NRAO\,150 is a very prominent millimeter to radio emitting quasar at redshift $z$=1.52 for which previous millimeter VLBI observations revealed a fast counterclockwise rotation of the innermost regions of the jet. 
Here we present new polarimetric multi-epoch VLBI-imaging observations of NRAO\,150 performed at 8, 15, 22, 43, and 86\,GHz with the Very Long Baseline Array (VLBA), and the Global Millimeter VLBI Array (GMVA) between 2006 and 2010. 
All new and previous observational evidence --i.e., spectral index maps, multi-epoch image cross-correlation, and low level of linear polarization degree in optically thin regions-- are consistent with an interpretation of the source behavior where the jet is seen at an extremely small angle to the line of sight, and the high frequency emitting regions in NRAO\,150 rotate at high speeds on the plane of the sky with respect to a reference point that does not need to be related to any particularly prominent jet feature.
The observed polarization angle distribution at 22, 43, and 86\,GHz during observing epochs with high polarization degree suggests that we have detected the toroidal component of the magnetic field threading the innermost jet plasma regions.
This is also consistent with the lower degree of polarization detected at progressively poorer angular resolutions, where the integrated polarization intensity produced by the toroidal field is explained by polarization cancellation inside the observing beam.
All this evidence is fully consistent with a kinematic scenario where the main kinematic and polarization properties of the 43\,GHz emitting structure of NRAO\,150 are explained by the internal rotation of such emission regions around the jet axis when the jet is seen almost face on.
A simplified model developed to fit helical trajectories to the observed kinematics of the 43\,GHz features fully supports this hypothesis.
This explains the kinematics of the innermost regions of the jet in NRAO\,150 in terms of internal jet rotation.} 
   
             

   \maketitle 
%

\section{Introduction} 
Very Long Baseline Interferometry (VLBI) has provided ultra-high-resolution observations of the innermost regions of jets in powerful blazars that reveal an increasing number of cases of  jets wobbling in the plane of the sky \citep{Mutel05, Agudo07, Marti11, Agudo12,Lister13}. 
The physical origin of blazar-jet curved structures \citep[e.g.,][]{Savolainen06} and helical paths of jet features \citep[e.g.,][]{Steffen95} is also thought to be related to the same phenomenon. 
However, there is no unique paradigm so far to explain this process.
Hence, the physical origin of blazar-jet wobbling --which is only observed in the innermost regions of blazar-jets and must therefore be closely related to the properties of the regions where the jet is formed, collimated, and accelerated-- is still far from being well understood.

Some of the most common scenarios proposed to explain this phenomenon involve the orbital motion of either the accretion disk or the jet nozzle, both induced by the presence of a companion supermassive compact-object \citep[e.g.,][]{Lister03, Stirling03}. 
These two scenarios may be useful to study the properties of the sources if they show periodic behaviors (e.g., jet precession), as reported for some well-known blazars such as 3C\,273 in \citet{Leppanen95} or 3C 345 in \citet{Lobanov05}. 
In contrast, there are different cases for which either it is not clear that the wobbling behavior is periodic \citep[e.g., in the case of BL Lac,][]{Mutel05}, or it is definitely not periodic at all \citep[e.g., OJ287 in][]{Agudo12}, hence suggesting that other kinds of jet instabilities may play a relevant role in the phenomenon \citep{Perucho12}.
To study the evolution of these instabilities it is fundamental to better understand the role that magnetic fields play in the innermost regions of jets in active galactic nuclei (AGN).
Although there are still uncertainties about the actual configuration of the magnetic field in such regions, current jet models and numerical simulations support the idea that the magnetic field is organized in a helical geometry along the inner jet and that the jet material traces this helical path following the field streamlines in the magnetically dominated jet region \citep{Vlahakis06, Marscher08_Nat,Mizuno12}. 
However, there is, so far, little direct observational evidence showing the jet plasma describing trajectories consistent with helical paths in radio loud AGN.
Looking for such evidence might be of great relevance to studying the jet wobbling phenomenon and the magnetic processes in the inner regions of relativistic jets in AGN. 

The source NRAO\,150 is an ideal object for conducting these studies. 
It is a powerful quasar at $z=1.52$ \citep{Acosta10} showing a significant misalignment $>100^{\circ}$ between the innermost jet regions observed with short millimeter VLBI (on submilliarcsecond scales), and those at larger distances from the central engine (on scales of tens of milliarcseconds and beyond). 
This pronounced misalignment is easily explained by invoking a slightly bent structure of the inner jet oriented within a very small angle to the line of sight \citep{Agudo07}.
One of the most interesting, and unusual, behaviors showed by the jet in NRAO\,150 is an extremely fast rotation of the inner jet structure (extended by $\sim0.5$\,mas) in the counterclockwise direction at an angular rate of up to $\approx$ 11$^{\circ}$/yr. 
This angular speed was estimated by \citet{Agudo07} by assuming that the brightest jet feature in most 43\,GHz VLBI images from 1997 to 2007 remained stationary, from which the remaining components were observed to move with superluminal speeds both in the radial and non-radial directions.

In this paper we present a new set of VLBI images at 8, 15, 22, 43, and 86\,GHz. 
The new data are used to follow the trajectories of the most prominent jet emission regions to make an updated characterization of the jet wobbling phenomenon in NRAO\,150. 
We also revisit the kinematic model previously proposed for NRAO\,150 in \citet{Agudo07} and we present an alternative scenario to explain the behavior of the source.
This new scenario is based on the idea that we are actually observing the internal rotation of the plasma around the jet axis.

In this study, we assume $H_0$ = 71 km s$^{-1}$ Mpc$^{-1}$,  $\Omega_m$= 0.27, and  $\Omega_{\Lambda}$= 0.73 \citep{Komatsu09}. Under these assumptions, the luminosity distance of NRAO 150 is dL = 11164.5 Mpc, 1 mas corresponds to 8.5 pc in the frame of the source, and an angular proper motion of 1 mas yr$^{-1}$ translates into a speed of 70.14 c. 


\section{Observations}

   \begin{figure}
   \centering  
   \includegraphics[height=21.cm]{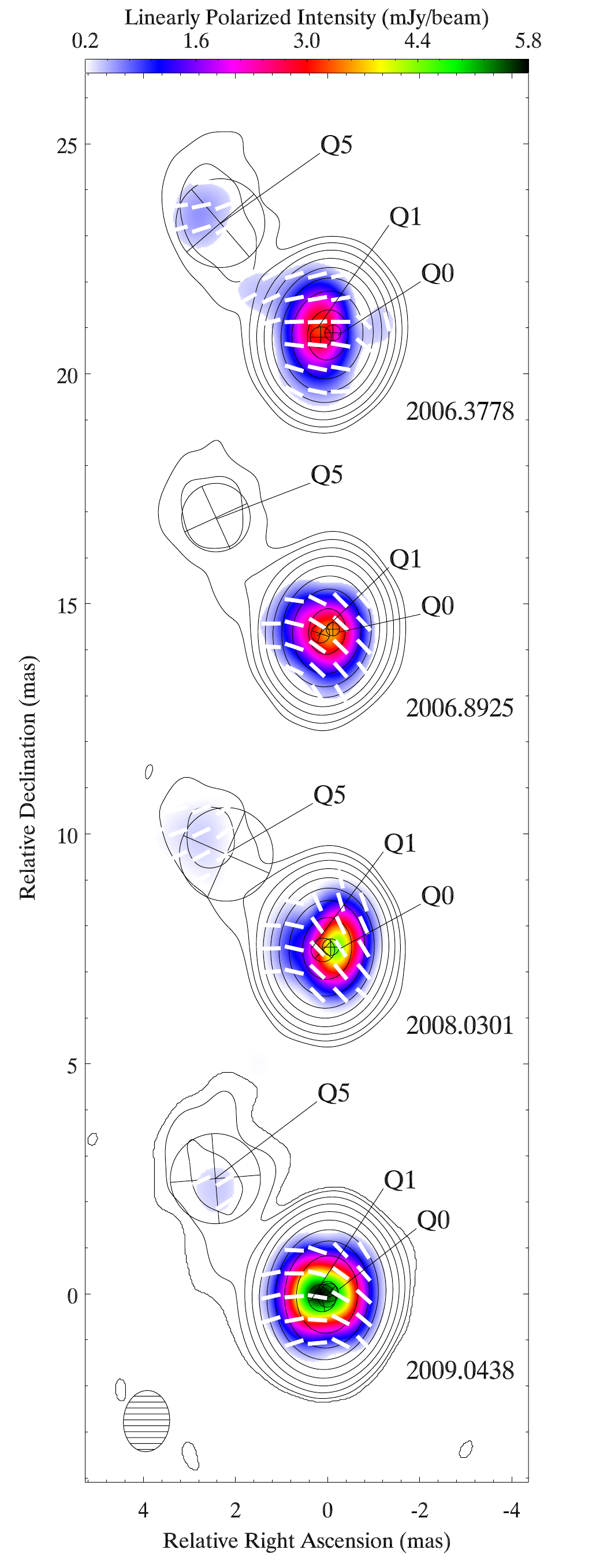} 
       \caption{Sequence of 8\,GHz VLBA images of NRAO\,150 from 2006 to 2009. The observing epoch has been labeled for every one of the images. Contours symbolize the observed total intensity, the color scale represents the linearly polarized intensity, and the short white sticks indicate the electric vector position angle distribution for every observing epoch. Total intensity contours are overlaid at 0.04 \%, 0.08\%, 0.2 \%, 0.48 \%, 1.15 \%, 2.75 \%, 6.57 \%, 15.72 \%, 37.61 \%, and 90 \% of the total intensity peak at 9.82 Jy beam$^{-1}$. A common convolving beam of FWHM 1.33 x 1.003 mas at -3.78$^{\circ}$ was used for all images and is shown in the lower-left corner.}
       \label{maps8}
   \end{figure}

   \begin{figure}
   \centering
      \includegraphics[height=23.cm]{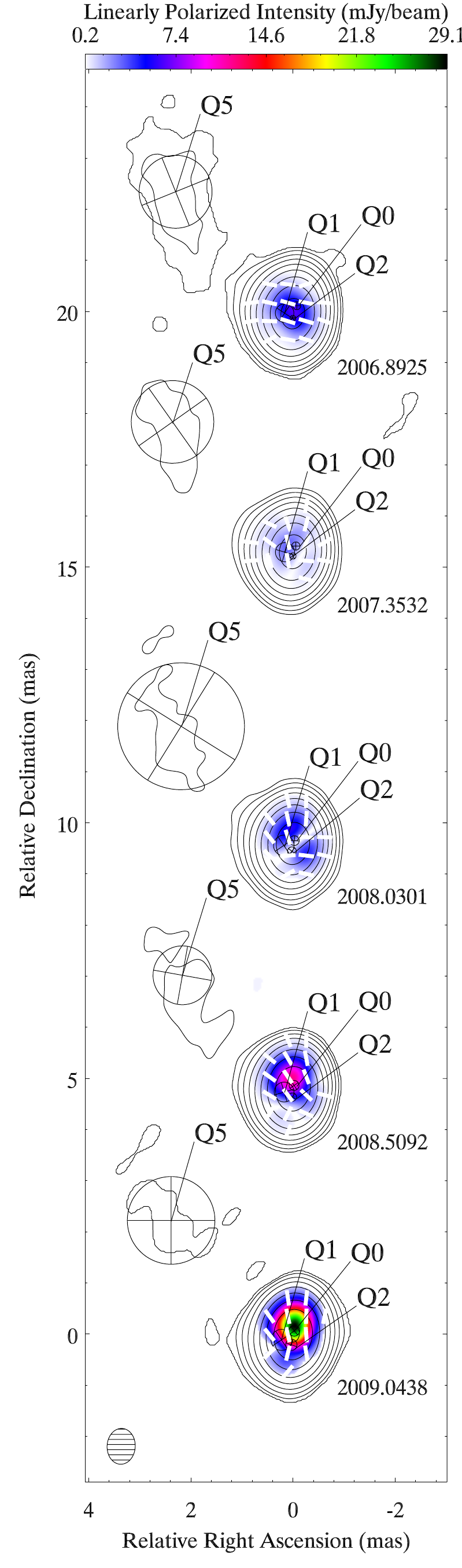}
       \caption{Same as Fig.~\ref{maps8}, but for the 15\,GHz VLBA images. Total intensity contours are overlaid at 0.04 \%, 0.11\%, 0.28 \%, 0.73 \%, 1.91 \%, 5.00 \%, 13.11 \%, 34.35 \%, and 90 \% of the total intensity peak at 11.71 Jy beam$^{-1}$. A common convolving beam of FWHM 0.7 x 0.55 mas at -3.00$^{\circ}$ was used for all images and is shown in the lower left-corner.}
       \label{maps15}
   \end{figure}

   \begin{figure}
   \centering
       \includegraphics[height=23.cm]{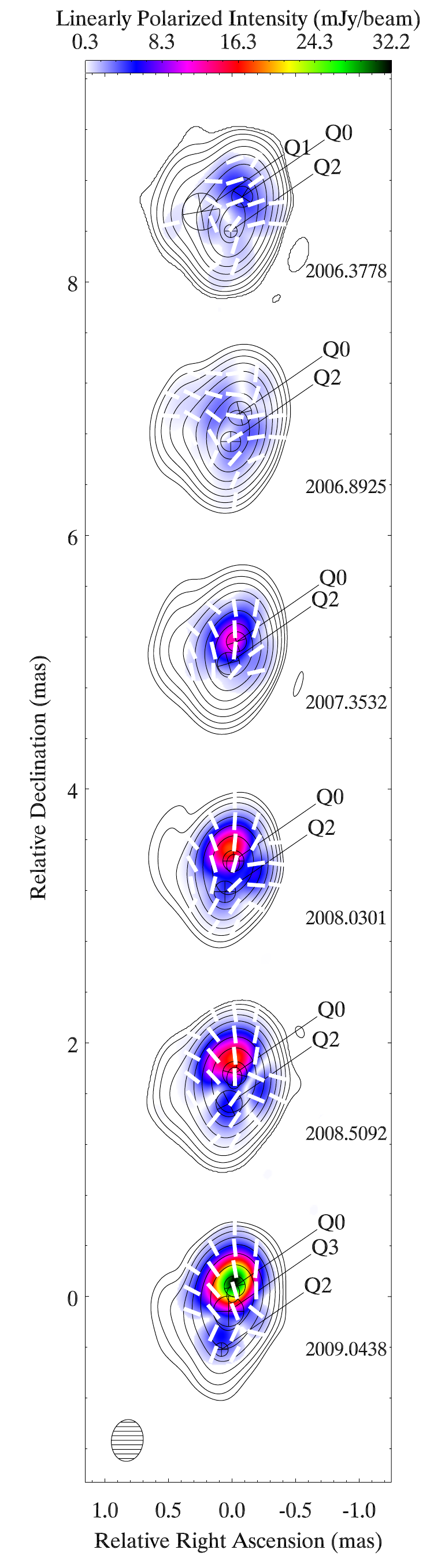}
       \caption{Same as Fig.~\ref{maps8}, but for the 22\,GHz VLBA images. Total intensity contours are overlaid at 0.40 \%, 0.87\%, 1.88 \%, 4.08 \%, 8.84 \%, 19.15 \%, 41.52 \%, and 90 \% of the total intensity peak at 6.06 Jy beam$^{-1}$. A common convolving beam of FWHM 0.33 x 0.25 mas at -6.40$^{\circ}$ was used for all images and is shown in the lower left-corner.}
       \label{maps22}
   \end{figure}

   \begin{figure}
   \centering
       \includegraphics[height=23.cm]{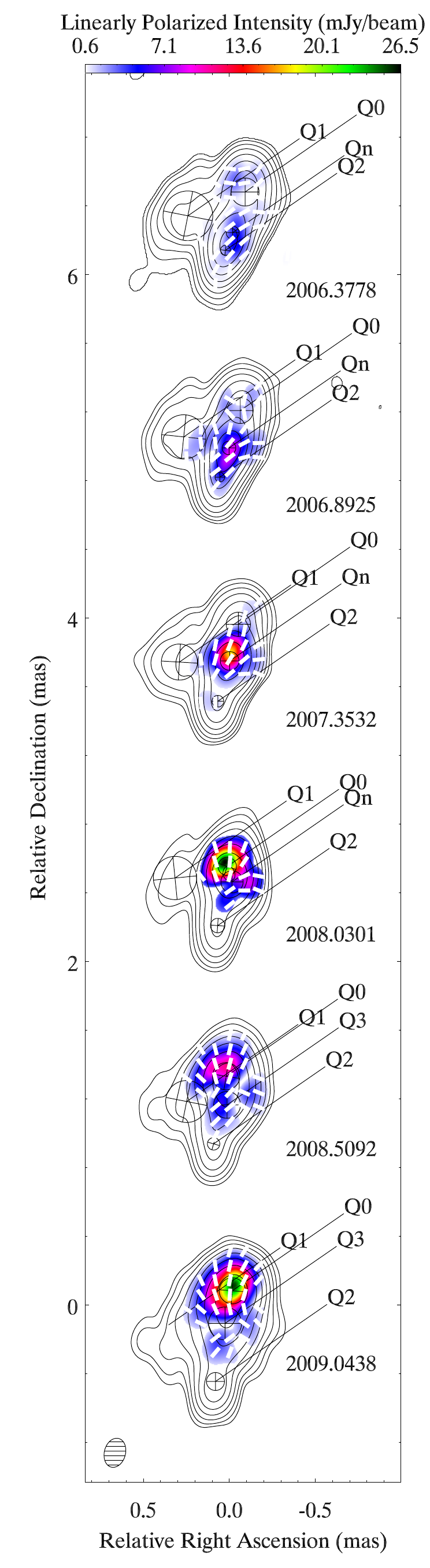}
       \caption{Same as Fig.~\ref{maps8}, but for the 43\,GHz VLBA images. Total intensity contours are overlaid at 0.19 \%, 0.41\%, 0.88 \%, 1.90 \%, 4.11 \%, 8.88 \%, 19.22 \%, 41.59 \%, and 90 \% of the total intensity peak at 3.61 Jy beam$^{-1}$. A common convolving beam of FWHM 0.17 x 0.12 mas at -14.85$^{\circ}$ was used for all images and is shown in the lower-left corner.}
       \label{maps43}
   \end{figure}

   \begin{figure}
   \centering
       \includegraphics[height=23.cm]{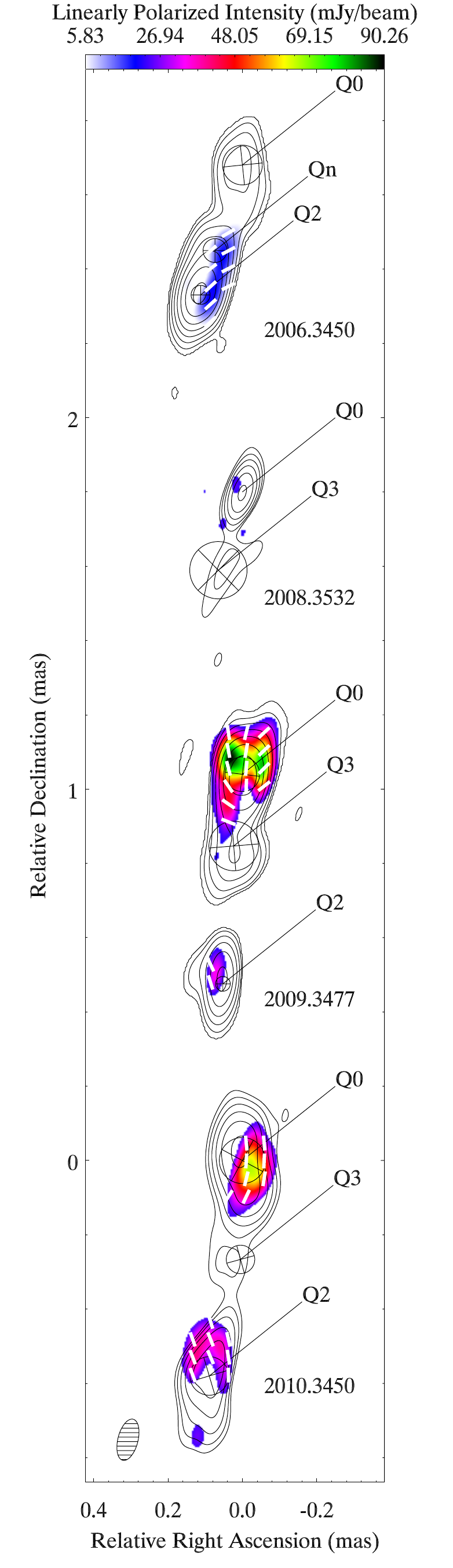}
       \caption{Same as Fig.~\ref{maps8}, but for the 86\,GHz VLBA images. Total intensity contours are overlaid at 2.78 \%, 4.97\%, 8.87 \%, 15.83 \%, 28.25 \%, 50.42 \%, and 90 \% of the total intensity peak at 1.47 Jy beam$^{-1}$. A common convolving beam of FWHM 0.11 x 0.05 mas at -14.34$^{\circ}$ was used for all images and is shown in the lower-left corner.}
       \label{maps86}
   \end{figure}

The Very Long Baseline Array (VLBA) and Global Millimeter VLBI Array (GMVA)\footnote{See http://www3.mpifr-bonn.mpg.de/div/vlbi/globalmm/} images of the total and linearly polarized intensity distributions of NRAO\,150 are presented in Figs.~\ref{maps8} to \ref{maps86}. The specific dates, frequencies, and interferometer used are listed in Table~\ref{Table_vlbi_data}. The observing sessions were performed with a recording rate of 512 Mb s$^{-1}$. The stations used during GMVA observation were: Effelsberg (100 m, MPIfR, Germany), Pico Veleta (30 m, IRAM, Spain), Plateau de Bure (six 15 m antennas working in phased-array mode, IRAM, France), Onsala (20 m, Sweden), Mets\"ahovi (14 m, Finland), and a subset of the Very Long Baseline Array (i.e., all the VLBA antennas equipped with 86\,GHz receivers, which are those at Brewster, Owens Valley, Mauna Kea, Pie Town, Kitt Peak, Fort Davis, Los Alamos, and North Liberty).

The initial phase and amplitude calibration of the total flux and polarimetric data was performed with the AIPS software following the standard procedure for polarimetric observations \citep[see, e.g.,][for details]{Agudo06, Gomez11}. Amplitude calibration errors lead to an estimated error in flux density of $\sim$15 \%. After the initial calibration, the data were edited, self-calibrated, and imaged both in total and polarized intensity with a combination of AIPS and DIFMAP software \citep{Shepherd97}.

Analysis of the instrumental polarization (the so-called D-terms) at 86\,GHz show that they remain reasonably stable across our observing epochs. As expected, the most sensitive stations such as Effelsberg, Pico Veleta, or Plateau de Bure, have small D-term values, usually below 5\%. The VLBA antennas show larger values (especially Pie Town and Brewster), occasionally exceeding 10-12\%. Calibration of the electric vector position angle (EVPA) at 8\footnote{For calibration of the 8\,GHz EVPA we did not use the data from NRAO\,150, but only that from the calibrators (BL\,Lac, DA193, OJ287). At this frequency, we detected a systematic rotation of the EVPA calibration from NRAO\,150 with respect to that from the remaining calibrators that may perhaps be explained by prominent extended emission in NRAO\,150 not detected in our VLBA images.}, 22, and 43\,GHz was obtained by comparison of the integrated polarization measured from VLBA images for NRAO\,150 and some calibrators (BL\,Lac, DA193, OJ287) with VLA observations at contemporaneous epochs. 
The same technique was employed to perform the absolute EVPA calibration of the 86\,GHz GMVA polarimetric images.
For this case, independent single-dish polarimetric observations of NRAO\,150 at 86\,GHz with the XPOL polarimeter \citep{Thum08} on the IRAM 30\,m Telescope data were obtained as a further support of IRAM to the GMVA.

At 15\,GHz, the EVPA calibration was performed by comparing the VLBA polarimetric images at the frequency provided by the MOJAVE team \citep{Lister09} for contemporaneous epochs with 15\,GHz observations in our program.
For our first two epochs at 15\,GHz (2006.89 and 2007.35), no MOJAVE polarization map reasonably close in time was available.  
For these two cases we made an alternative calibration based on the stability of the instrumental polarization terms (D-terms) at the VLBA \citep{Gomez02_b}. 
The EVPA calibration obtained was consistent in all cases with the D-term values across observing epochs.
Estimated errors in the orientation of EVPA vary with observing epoch and frequency, but lie in the range between $5^{\circ}$ and $10^{\circ}$.

\begin{table*}
\caption{VLBI observations of NRAO 150.}   

\label{Table_vlbi_data}      


\centering                          
\begin{tabular}{l l l  }    
\hline 
 Date  &  Frequencies (GHz) &  Interferometer   \\ 
                 
\hline   
\hline
2006-May-19 (2006.37)  &   8, 22, 43 & VLBA  \\
\hline
\hline
2006-May-07 (2006.43)  &   86 & GMVA \\
\hline	
\hline
2006-November-23 (2006.89) &   8, 15, 22, 43 & VLBA \\

\hline
\hline
2007-May-10 (2007.35) &   15, 22, 43 & VLBA  \\

\hline	
\hline
2008-January-12 (2008.03) &   8, 15, 22, 43 & VLBA  \\

\hline	
\hline
2008-May-08 (2008.43) &   86 & GMVA  \\
\hline	
\hline
2008-July-05 (2008.50)    &   15, 22, 43 & VLBA  \\

\hline	
\hline
2009-January-17 (2009.04) &   8, 15, 22, 43 & VLBA  \\

\hline	
\hline
2009-May-8 (2009.43) &   86 & GMVA \\
\hline	
\hline
2010-May-6 (2010.43)	&   86 & GMVA  \\	 
\hline 		
\end{tabular}
\end{table*}

\section{Data analysis}

At 8\,GHz and 15\,GHz the NRAO\,150 brightness distribution is dominated by a compact emission region.
A weaker jet-like extended structure towards the northeast is also systematically detected (see Figs.~\ref{maps8} and \ref{maps15}). The 22\,GHz and 43\,GHz VLBA images (see Figs.~\ref{maps22} and \ref{maps43}) reveal a much richer structure of the innermost regions of NRAO\,150, as well as a clearer variable emission structure.

At these two frequencies, we identify a dominant north-south total intensity structure at the top of an extended region towards the east. 
As observed in total flux, the linear polarization distribution is also variable, showing a peak in the northern region at the 22\,GHz and 43\,GHz images in 2009.04.
This highly polarized structure is also displayed by the 86\,GHz images in 2009 and 2010 (see Fig.~\ref{maps86})

\subsection{Model fits and identification of prominent emission features}

To obtain a simple and easily treatable representation of the total brightness distribution of the jet in NRAO\,150, we employed a set of circular Gaussian emission components fitted to the actual visibility data in the UV plane. 
The minimum number of Gaussian components describing all jet regions and at close positions during contiguous epochs was employed at every observing frequency.
The modelfit task inside the DIFMAP package was used for this. 
Tables~\ref{Table_2} to \ref{Table_6} list the resulting model-fit parameters of every one of the fitted emission features at every observing epoch and frequency. 

For the naming of Gaussian emission components we used essentially the same nomenclature as in \citet{Agudo07}, although the northern component is named Q0 here, instead of the Core. 
Within our new six observing epochs at 43\,GHz we identify five features: Q0, Q1, Q2, Q3, and Qn.
Features Q0, Q2, and Qn align along the north-south direction of the source, whereas Q1 corresponds to the easternmost region of the brightness distribution observed at 22\,GHz and 43\,GHz. 
At the two lower observing frequencies (8\,GHz and 15\,GHz) we also identify an extended component (Q5) as part of the large-scale jet oriented towards the northeast. Model fit of the last epoch (2010.43) at 86\,GHz allows for the inclusion of an extra component in the region near Q0, but for consistency with previous epochs/frequencies only one component associated with Q0 has been considered.

Beginning on our first new observing epoch in 2006.37 we report the dominance of a new emission feature (Qn), which has a drastically different nature than any other component previously detected in NRAO\,150 at 43\,GHz.
Unlike Q0, Q1, Q2, and Q3, Qn moved describing an essentially straight motion (see below); it started as a prominent flare in total flux from the beginning of 2006 to mid 2008, and moved at a fast superluminal speed of $6.3\pm1.1\,c$ towards the north (considering Q0 as the stationary kinematic center).
Since Qn dominated the jet region surrounding its position along its path to the north, it was not possible to discern the Q3 component during some observing epochs (from 2006.38 to 2008.03).

The uncertainties of the model-fit parameters presented in Tables~\ref{Table_2} to \ref{Table_6} were computed by following the method described in \citet{Jorstad05}. 
The prescriptions by \citet{Jorstad05} were defined for the uncertainties in 43\,GHz-image model fits only.
We have extended and modified the method to consider the uncertainties of the model-fit parameters at other observing frequencies.
The \citet{Jorstad05} estimated uncertainties are based on the total flux and size of the fitted emission features, and take into account that bright and compact features are more accurately fitted than those weaker and more extended.
In particular, we used the following criteria for the estimates of the uncertainties of our model-fit parameters:
(1) for bright and compact features with flux of knot $\geq100$\,rms noise level and size $\leq2/3$ beam, the uncertainties in flux density were taken as $\sim1\,\%$, 1/10 of beam size in position, and  $1\%$ in size; 
(2) for bright and more extended features with flux $\geq100$\,rms noise level and size $\geq2/3$ beam, the uncertainties in flux density were estimated as $\sim3\%$, 1/5 of the beam size in position, and $10\%$ in size;
(3) for weak and non-extended features with flux $<$ 100\,rms noise level and size $<$ the beam size, the uncertainties in flux density were estimated as $\sim10\%$, 1/5 of the beam size in position, and $5\%$ in size; and
(4) for weak and extended features with flux $<$ 100\,rms noise level and size $>$ the beam size, we applied uncertainties $\sim10\%$ in flux density, 1/2 of the beam size of the fitted feature in position, and $10\%$ in size.
We also added in quadrature an intrinsic flux error of 15\% due to the uncertainties in the amplitude calibration.

In Tables~\ref{Table_2} to \ref{Table_6} we present the fitted parameters of the jet features for each frequency and observing epoch. 
The columns of these tables are as follows: (1) observing epoch, (2) name of emission component, (3) flux density in Janskys, (4) relative right ascension in milliarcseconds, (5) relative declination in milliarcseconds, (6) angular size in milliarcseconds, (7) degree of polarization, (8) EVPA.



\subsection{Image alignment}

During the self-phase-calibration of VLBI images the information of the absolute position of the source in the sky is lost. 
Therefore, a basic but critical requirement for comparing and/or combining VLBI images at different epochs or observing frequencies is to properly align such images.
Two main methods are typically employed for this purpose. 

The first consists in the cross identification of a particular emission feature (whose position is defined by its Gaussian model fit) in the two different images to analyze. 
For this, a compact optically-thin region detectable in both images is needed.
Optically thick regions are affected by synchrotron self-absorption, which changes their position along the jet as a function of emitting frequency. 

The second method consists in the two-dimensional (2D) cross-correlation of the optically thin emission distribution of the source \citep{Walker_2000, Croke08}. During the cross-correlation process, one of the images is shifted in the $\Delta$x and the $\Delta$y axes (where x and y are the right ascension and declination coordinates in the plane of the sky, respectively) and the cross-correlation coefficient at each position shift ($\Delta$x, $\Delta$y) is calculated. 
The best position shift corresponds to the maximum cross-correlation coefficient. 
Following \citet{Croke08}, the 2D cross-correlation coefficient was calculated as

\begin{equation}
r_{xy}={{ \sum _{i=1}^n \sum _{j=1}^n (I_{\nu1,ij} - \overline{I_{\nu1}}) (I_{\nu2,ij} - \overline{I_{\nu2}} ) }  \over {\sqrt { \sum _{i=1}^n \sum _{j=1}^n (I_{\nu1,ij} - \overline{I_{\nu1}})^2  \sum _{i=1}^n \sum _{j=1}^n (I_{\nu2,ij} - \overline{I_{\nu2}} )^2 }}} ,
\label{ec1}
   \end{equation}
where n is the number of pixels in both the x and y axes, i and j are the indexes of pixels in each of these two axes, $I_{\nu1,ij}$ is the intensity of the unshifted image at position (i, j), and $I_{\nu2,ij} $ is the intensity of the shifted image at position (i, j); $\overline{I_{\nu1}}$ and $\overline{I_{\nu2}}$ are the average intensities over the region analyzed at frequencies $\nu1$ and $\nu2$, respectively.

This second method has the advantage that it takes into account all optically thin emission regions on the entire source, not just a particular isolated compact feature, and it is extremely useful for smooth/weak sources where it is not possible to clearly distinguish the position of emission features to align the different images \citep[e.g.,][]{OSullivan09}.

The particular case of NRAO\,150 is especially difficult, since we could not find optically thin components detected at all our observing frequencies. 
Therefore, we had to compare the results from both methods outlined above and then use the most robust results; see below for a detailed description of each case. 
In all cases we used the restoring beam size of the lower frequency image for every pair of images to be compared.

To align images at 8\,GHz and 15\,GHz we used a pixel size of 0.03\,mas and a beam size of 1.33 x 1.00\,mas  at position angle (PA)$=-3.78^{\circ}$ at both frequencies.
In all images at these two low frequencies it is possible to distinguish the optically thin region Q5 (Figs.~\ref{maps8} and \ref{maps15}) that was initially used to align the images.
However, this method gave us less reliable results than with the 2D cross-correlation method. 
This is expected, since Q5 is an extended region and hence the uncertainty in its position is rather large. 
We therefore employed method 2 to register the 8\,GHz and 15\,GHz images, which allowed us to correlate the positions of the entire optically thin region in the 8\,GHz and 15\,GHz jets, not just the position of a particular (extended) jet feature. 

To align the 15\,GHz and 22\,GHz images we employed a pixel size of 0.03\,mas and a beam size of 0.7 x 0.55\,mas at position angle (PA)$-3.00^{\circ}$.
At 22\,GHz, the Q5 emission feature is over-resolved by the VLBA, and therefore it is not possible to use this component to align the 15\,GHz and 22\,GHz pairs of images.
In addition, using the position of Q0 or Q2 features does not provide consistent results. 
Therefore, we used the 2D cross-correlation method, which gave considerably more consistent results than using method 1.

For the comparison of the 22\,GHz and 43\,GHz images, we used a common pixel size of 0.01\,mas, a beam size of 0.33x0.25\, mas, and PA$=-6.40^{\circ}$. For the case of these high-frequency images, image alignment based on the cross identification of a particular emission feature, method 1, also gave completely inconsistent results when using either Q0 or Q2 reference positions for alignment. Cross identification using Q0 leads to a growing optical depth towards the southern jet regions, in which Q0 is optically thin while Q2 becomes thick; using Q2 gives the opposite result, with increasing opacity towards the north, Q0 becomes optically thick and Q2 thin. We also note that the cross identification by using Q0 or Q2 involves image shifts of approximately 1/10 of the beam size or smaller, yet this implies significant changes in the opacity of the jet, with absolute variations of about 0.5-1 in the spectral index.

We therefore used 2D cross-correlation to align the images at 22\,GHz and 43\,GHz (Fig.~\ref{sp22_43}). Final registering of the images is obtained after masking the optically thick regions resulting from the initial cross-correlation. The alignment gave us a rather homogeneous spectral index distribution compared with that from method 1. We note that the spectral index images at these higher frequencies are subject to a significant degree of uncertainty due to the difficulties in obtaining a reliable image alignment mentioned previously.
 
For the 43\,GHz and 86\,GHz image comparison, we have used a pixel size of 0.005\,mas and a beam size of 0.17 x 0.12\,mas and PA$=-14.85^{\circ}$. 
We set a maximum allowed time between the 43\,GHz and 86\,GHz images for our spectral index maps of four months.
For the case of the 43\,GHz to 86\,GHz spectral index maps, we obtained similar results either employing cross identification with Q0 or Q2 emission features, or using 2D cross-correlation for image alignment showing almost all source optically thin emission.
We therefore used 2D cross-correlation to align the images, but the quality of the 86\,GHz image during 2008.35 is rather poor because during observation we had bad weather in several important antennas, which required editing a significant amount of data. For this reason, the brightness distribution is not coincident between 43\,GHz and 86\,GHz. In this case, method 1 gives us a more confident result, since the Q0 feature is compact, well defined, and optically thin at both 43\,GHz and 86\,GHz. So to align the second map showed in Fig.~\ref{sp43_86}, we used cross identification based on Q0 feature for image alignment between these two observing frequencies.

\subsubsection{Rotation between epochs}

The source NRAO\,150 shows an extremely fast counterclockwise rotation of the inner source structure \citep{Agudo07}.
This rotation was measured considering Q0 as the stationary core. 
However, the spectral index maps (see Figs.~\ref{sp8_15} to \ref{sp43_86}) do not reveal any obvious region that could be considered the core from our new data.
Therefore, we tried to measure the rotation of the source without any previous assumption about the kinematic center of the system.
To measure the rotation in a representative time span, we used the 43\,GHz image sequence published by \cite{Agudo07} and the new six observing epochs at 43\,GHz presented in this work.

For that, first we employed the 2D cross-correlation method to align 43\,GHz images at successive epochs. Second, we used the 2D cross-correlation method
between epochs, rotating one of the images instead of applying a shift in right ascension and declination. 
To avoid problems with the corners of the matrix during the process -- for image combination we used images set as square matrixes-- we used a circular mask that flags the edges of the matrix, while always conserving the entire source structure. 
The expression involved in the computation of the rotation cross-correlation index is

\begin{equation}
r_{xy}^{rot}={{ \sum _{i=1}^n \sum _{j=1}^n (M_{1,ij} - \overline{M_{1}}) (M^{rot}_{2,ij} - \overline{M_{2}} ) }  \over {\sqrt { \sum _{i=1}^n \sum _{j=1}^n (M_{1,ij} - \overline{M_{1}})^2  \sum _{i=1}^n \sum _{j=1}^n (M^{rot}_{2,ij} - \overline{M_{2}} )^2 }}},
\label{ec2}
 \end{equation}
where M represents the masked intensity of the images. 
As in equation (\ref{ec1}), $M_{1,ij}$ and $M^{rot}_{2,ij}$ are the masked intensities without rotation and rotated at two different epochs, respectively. $\overline{M_{1}}$ and $\overline{M_{2}}$ are the averaged mask intensities over the analyzed region at those epochs.
We rotated the images around the center of the images with an increment of 1$^{\circ}$ until a maximum value of 30$^{\circ}$ in clockwise and counterclockwise sense. 

Therefore, the maximum cross-correlation coefficient gives us the rotation between two 43\,GHz images taking into account the intensity distribution of the entire source in both images. 
It is important to note that if the intensity distribution of one of the images has a very prominent peak, the 2D cross-correlation tends to align its most prominent peaks without giving much relevance to the remaining intensity distribution of the remaining source regions. 
For this reason, it is not practical to compare observing epochs involving prominent changes in the intensity distribution. 
In addition, if we compare images in consecutive epochs with little structural changes between them, detecting a rotation between images is very difficult (sometimes impossible) owing to the relatively slow source evolution in time scales of a few months. 
Thus, we have compared epochs between the maximum possible time span that does not include significant changes in the intensity distribution of the images. 
The variations in the intensity distribution are not constant during all epochs observed, so the time interval used was different in different moments. The minimum time range used to compare epochs was 0.5 year and the maximum was 1.9 year. 
Before 2000.62 the source does not show asymmetry that is sensitive enough to be measured with our method, so we do not include epochs before this time in this analysis.

We find a total rotation between 2000.62 and 2009.04 observing epochs of 19$^{\circ}$ in the counterclockwise direction.  
The rotation seems to be considerably faster between epochs 2002.48 and 2004.97 compared with the remaining time ranges of our monitoring (see, e.g., Fig. 2 in Agudo et al. 2007).
After this date the rotation is more difficult to measure because of the poorer observing sampling and also because of the more  prominent change in source structure in observing epochs before $\sim2005$. 
Despite these difficulties, we measured a counterclockwise rotation of the jet structure of 4$^{\circ}$ from 2006 to 2009.

We stress that these results confirm --in an independent way compared to previous work-- the rotation of NRAO\,150's structure in the innermost regions mapped by 43\,GHz VLBI, and that the detection of this rotation does not depend on the reference point assumed to model the kinematics of the source.

\subsection{Spectral analysis}

The core region in relativistic jets is typically observed partially optically thick (i.e., showing positive spectral index $\alpha$, with $S_\nu\propto\nu^{\alpha}$, being $S_\nu$  the source flux density at observing frequency $\nu$) while the remaining jet regions are usually optically thin ( $\alpha$ $<0$). Therefore, maps of the spectral index along the jet help us to identify regions with different properties. For the spectral index studies presented in this paper, we compared images at two adjacent frequencies, both of them aligned following Sect.~3.2. A Gaussian uv-taper is used in the higher frequency data in order to obtain a similar (u,v) coverage, followed by a convolution with the same beam as the one at the lower frequency. Despite this, the intrinsic uneven (u,v) coverage at different frequencies may introduce some uncertainties in the final spectral index maps, but these are probably negligible when compared to those related to the image alignment.

\begin{figure}
   \centering
   \includegraphics[height=19.cm]{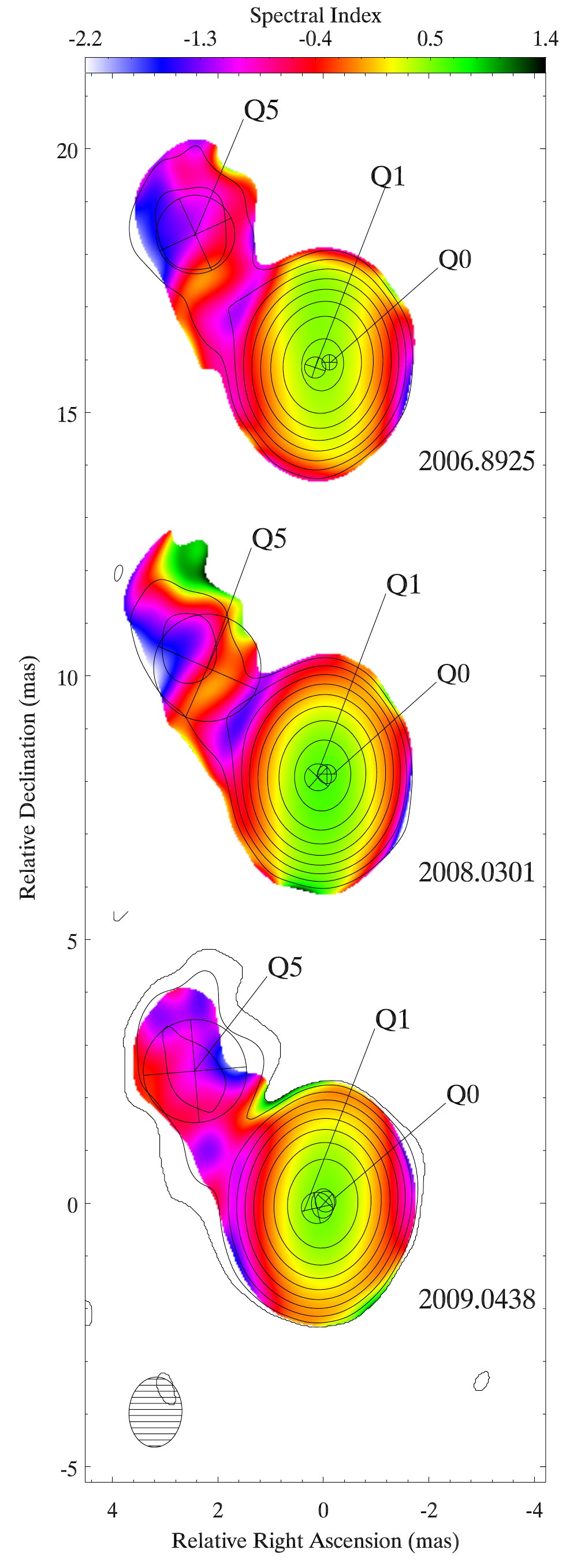}
       \caption{8.4\,GHz to 15.4\,GHz spectral index distributions overlaid with 8.4\,GHz total intensity contours. Images at the two observing frequencies were convolved with a restoring beam with FWHM 1.33 x 1.00\,mas and PA = -3.78$^{\circ}$.}
       \label{sp8_15}
   \end{figure}

In the spectral index maps at the lowest frequencies considered in this paper,  8.4\,GHz and 15.4\,GHz (Fig.~\ref{sp8_15}), the innermost region of the jet is clearly optically thick, with spectral index decreasing towards the northeastern jet regions, where the Q5 component is located and the jet becomes optically thin.

\begin{figure}
   \centering
   \includegraphics[height=23.cm]{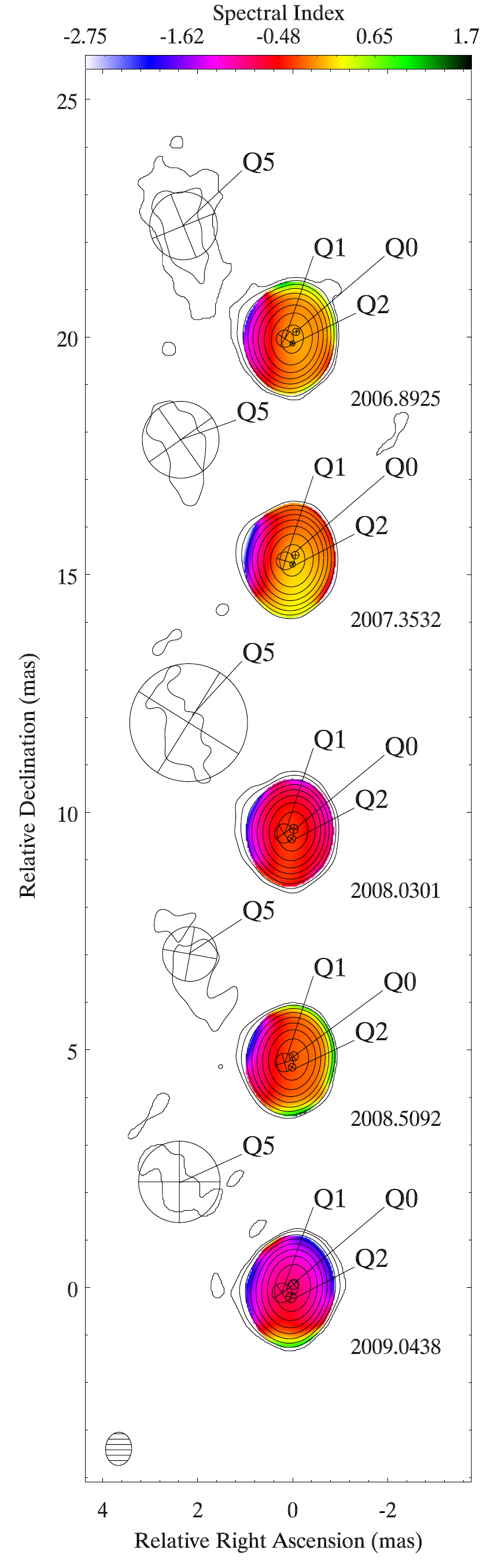}
       \caption{Same as Fig.~\ref{sp8_15}, for the 15.4\,GHz to 22.2\,GHz spectral index distributions overlaid with 15.4\,GHz total intensity contours. Images at the two observing frequencies were convolved with a restoring beam with FWHM 0.7 x 0.55 mas and PA = -3 $^{\circ}$.}
       \label{sp15_22}
   \end{figure}

In the spectral index maps at 15.4\,GHz to 22.2\,GHz (see Fig.~\ref{sp15_22}) we begin to observe some optically thin regions in the central region of NRAO\,150. 
In the first two epochs in Fig.~\ref{sp15_22}, the source shows a spectral index consistent with a flat spectrum, while the last three epochs show a more optically thin spectrum in most of the central region of the source.

\begin{figure}
   \centering
   \includegraphics[height=23.cm]{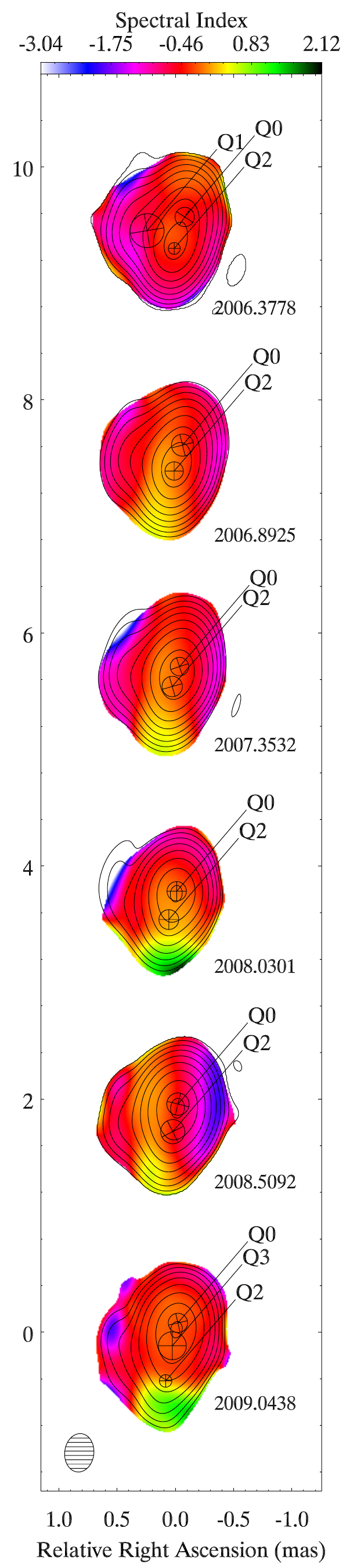}
       \caption{Same as Fig.~\ref{sp8_15}, for the 22.2\,GHz to 43.2\,GHz spectral index distributions overlaid with 22.2\,GHz total intensity contours. Images at the two observing frequencies were convolved with a restoring beam with FWHM 0.33 x 0.25 mas and PA = -6.40 $^{\circ}$.}
       \label{sp22_43}
   \end{figure}

At higher frequencies, the 22\,GHz to 43\,GHz spectral index maps (see Fig.~\ref{sp22_43}) show that the emission is mostly optically thin with a tendency towards an optically thick spectrum in the southern region. This optically thick region is not coincident with any particularly prominent component, except perhaps for epoch 2009.04, in which component Q2 shows a flat-to-optically thin spectrum. 

\begin{figure}
   \centering
   \includegraphics[height=14.cm]{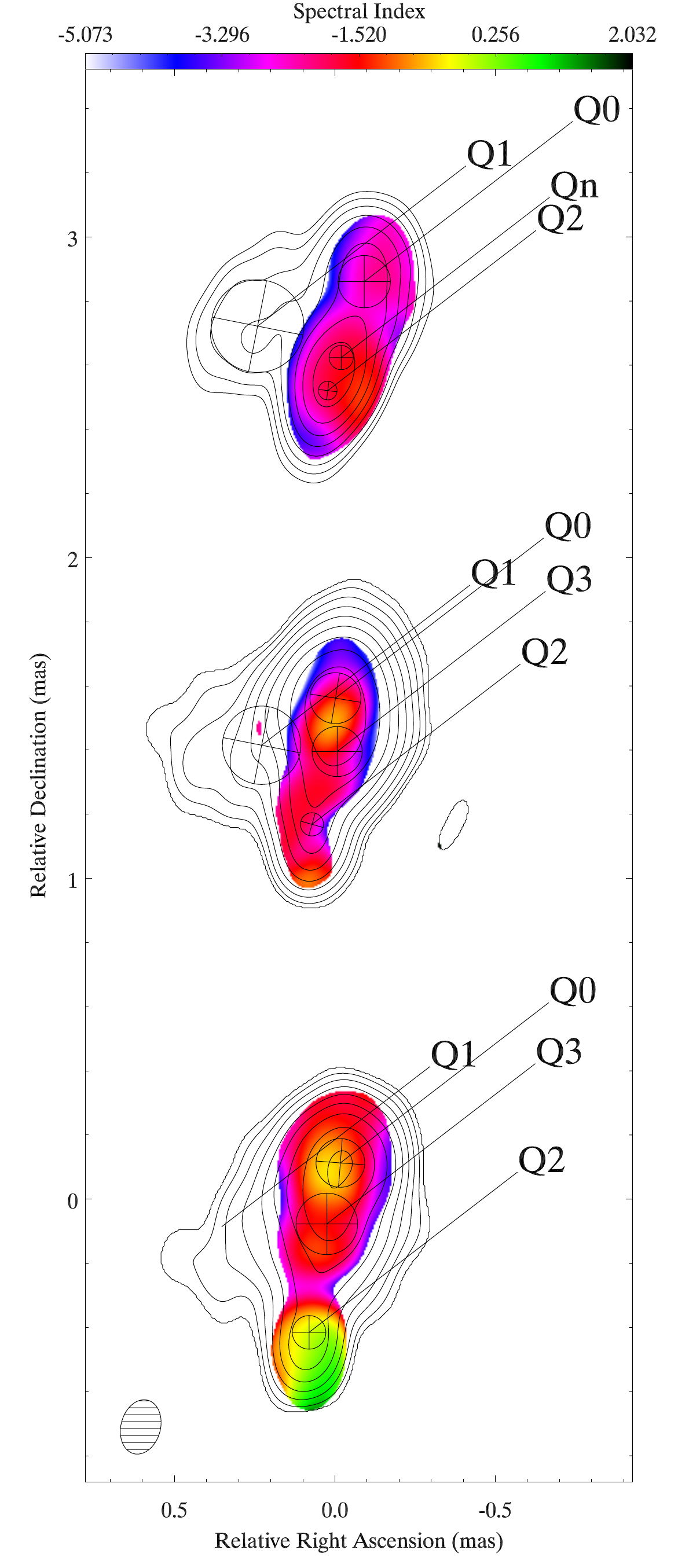}
       \caption{Same as Fig.~\ref{sp8_15}, for the 43.2\,GHz to 86.2\,GHz spectral index distributions overlaid with 43.2\,GHz total intensity contours. Images at the two observing frequencies were convolved with a restoring beam with FWHM 0.17 x 0.123 mas and PA = -14.85 $^{\circ}$. The first map was built with data at 43.2\,GHz and 86.2\,GHz taken during 2006.37 and 2006.43, the second with data taken during 2008.50 and 2008.43, and the last during 2009.04 and 2009.43. }
       \label{sp43_86}
   \end{figure}

Finally, the 43\,GHz to 86\,GHz spectral index map (see Fig.~\ref{sp43_86}) shows all the emission components (except for Q2 at the beginning of 2009) as clearly optically thin regions. 

\begin{figure}
   \centering
       \includegraphics[width=9.cm]{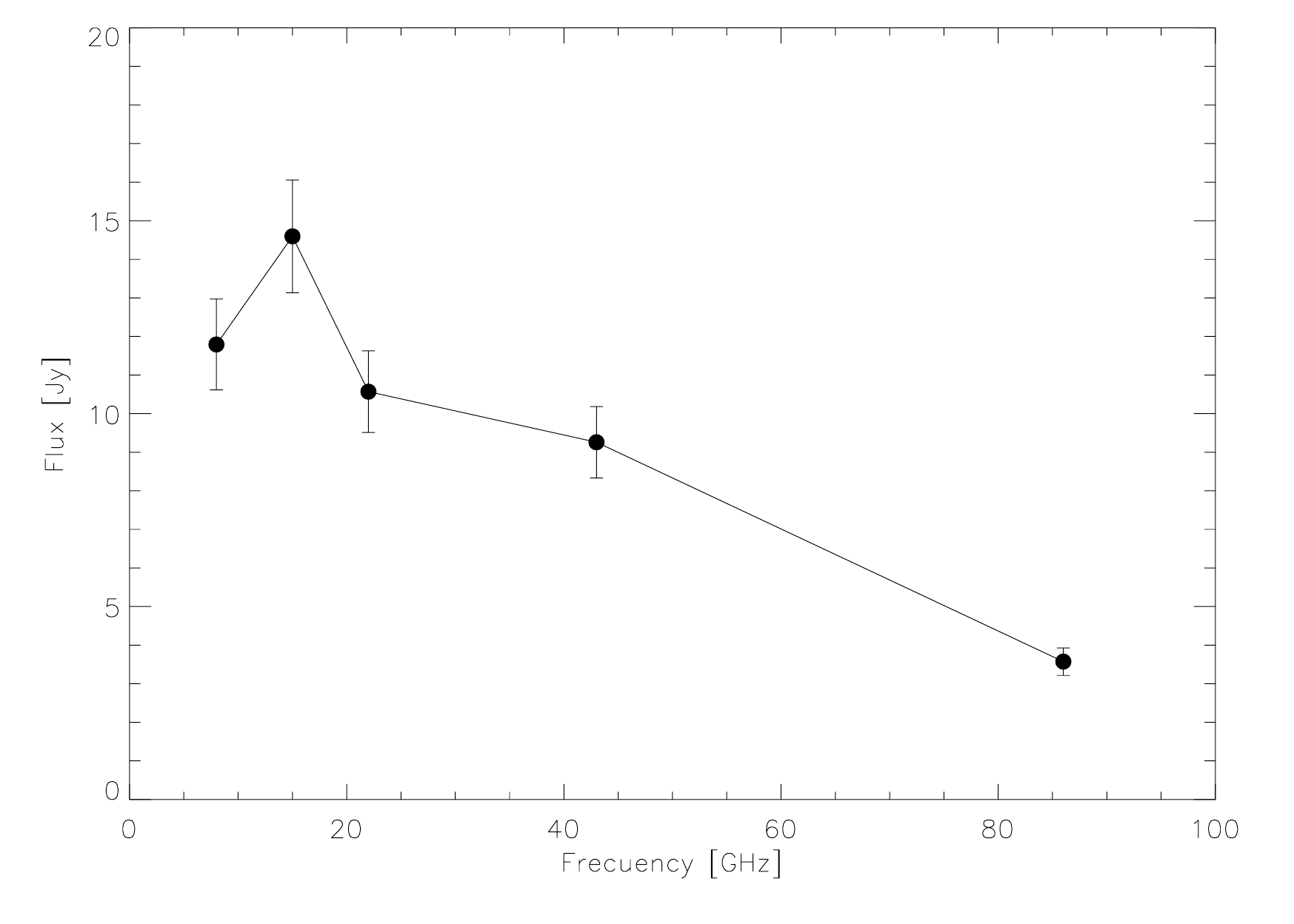}
       \caption{8.4\,GHz to 86.2\,GHz spectrum for NRAO\,150 built from the integrated total flux images on epochs 2009.04 at 8.4, 15.4, 22.2, and 43.2\,GHz and epoch 2009.34 at 86.2\,GHz.}
       \label{sp_epocaF}
   \end{figure}

In Fig.~\ref{sp_epocaF} we show the spectrum of NRAO150 across all of our observing frequencies, obtained for epoch 2009.04 (except at 86\,GHz, for which we used the GMVA image on 2009.34). The emission at 8\,GHz is optically thick while the peak of the NRAO\,150 spectrum is located at $\sim15$\,GHz. At higher frequencies the emission becomes optically thin.

In summary,  the source shows a homogeneous behavior over the central region (the innermost $\sim0.5$\,mas as observed on our 22, 43, and 86\,GHz images) showing that, unlike in other blazars \citep[e.g.,][]{ Beckert02, Croke10} there is not substantial evidence that any of the observed features is systematically optically thick as expected for the core.

\subsection{Structure of linear polarization intensity and EVPA distributions}

Figures~\ref{maps8} to \ref{maps86} show the polarized intensity and electric vector position angle distribution (represented by sticks) at 8.4, 15.4, 22.2, 43.2, and 86.2\,GHz, respectively, whereas Figs.~\ref{pol_8} to \ref{pol_43} represent the linear polarization degree at 8.4, 15.4, 22.2, and 43.2\,GHz.
The latest set of images show that, except for the case of 8.4\,GHz images, the linear polarization degree increases with time from 2006 to 2009 at the northern emission region observed at the 43\,GHz region where the Q0 component is located. 
The linear polarization degree provides relevant information about the level of order (or turbulence) of magnetic fields along the line of sight in synchrotron sources.
In particular, the region around Q0 in epochs progressively closer to 2009 seems to be dominated by a more ordered field. 

The 43\,GHz images in Fig.~\ref{maps43} clearly show the fast motion of the Qn component from south to north over the three years of observations of our new observing program.
It is straightforward to see that the peak of the 43\,GHz linear polarization distribution coincides with the position of the Qn component. 
Although in epochs 2008.50 and 2009.04 we cannot identify separately Q0 and Qn, it is likely that Qn continues contributing significantly to the 43\,GHz polarization emission in the northern region of the source in these two epochs. 
In addition, the EVPA ($\chi$) orientation in the region near Qn is always parallel to the direction propagation of this emission feature, hence suggesting that the magnetic field in this region is perpendicular to the path followed by Qn. 
Therefore, we suggest that the Qn emission feature is a propagating perturbation in the jet that compresses and orders the magnetic field, increasing the polarization level and aligning the EVPA in the direction of its propagation.

\subsubsection{Evidence for a toroidal component of magnetic field}

Toroidal fields in relativistic jets are widely accepted as a natural consequence of the jet formation process \citep{Meier01_Science,Hardee05,McKinney09}, and are indirectly suggested by Faraday-rotation observations of the innermost regions of relativistic jets with VLBI \citep{Asada02, Gabuzda05, Gomez08, Croke10, Gomez11}. However, only one previous work \citep{Zamanina13} claims to have directly observed an ordered helical magnetic field in the jet of quasar 3C~454.3, although not for a complete circle and for spatial scales on the order of tens of parsecs.

\begin{figure*}
       \includegraphics[width=18cm]{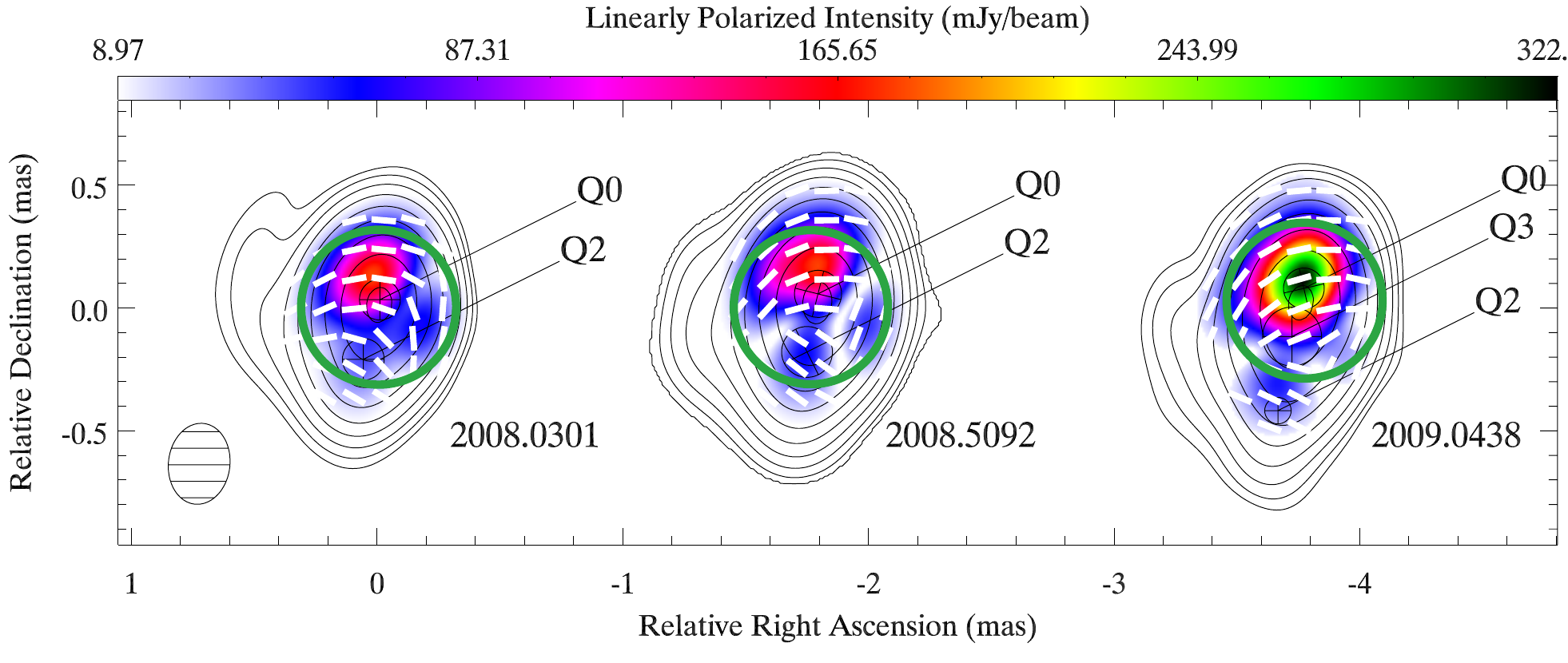}
       \includegraphics[width=18cm]{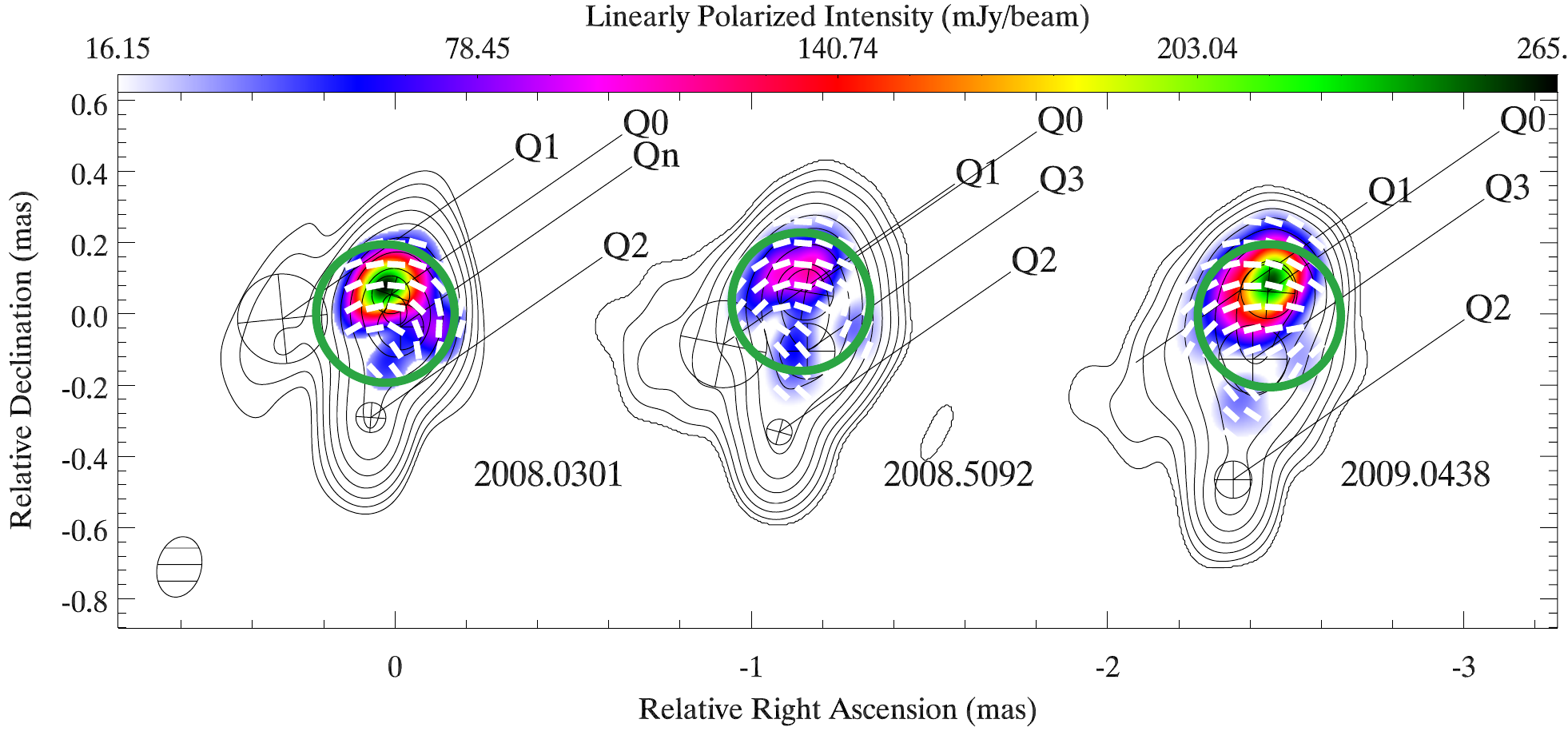}
       \caption{22\,GHz (top) and 43\,GHz (bottom) magnetic vector position angle distributions (symbolized by the short white bars) for the last three observing epochs in our new program as projected on the plane of the sky. The green line represents the toroidal component of the magnetic field that could produce the observed magnetic vector distribution. }
       \label{magnet_vector}
   \end{figure*}

In the present paper, we report an intriguing tendency of the magnetic vector position angle (that may be identified with the projected magnetic field of the jet in the plane of the sky) to distribute with a configuration that is fully consistent with a circular geometry (Fig.~\ref{magnet_vector}).
This is especially easy to observe in our last three observing epochs at both 22\,GHz and 43\,GHz, in which the linear polarization degree is larger and the magnetic field is therefore most probably dominated by a better ordered component than in previous observing epochs. 
This exceptional polarization structure can only be explained in a consistent way by invoking the toroidal component of the magnetic field as seen across its cross section when the jet is observed essentially face on. This structure is observed in the innermost regions of the jet observed at higher frequencies.

In addition, at 86\,GHz, the polarization structure is fully consistent with that in our 43\,GHz images, hence confirming the good capabilities of the GMVA for both 86\,GHz total flux and linear polarization imaging \citep[see also][and references therein]{Marti12}. 
In the 2006.43 (see Fig.~\ref{maps86}) 86\,GHz image, as well as at 43\,GHz, the peaks in total intensity and polarization are in the Q2 southern region. 
In contrast, at epoch 2009.04 the 86\,GHz total intensity and polarization peaks are in the northern region where, based on the 43\,GHz image sequence analysis, we expect the Qn jet feature, that we cannot separate from Q0 even in the higher resolution 86\,GHz map. 

In summary, observations at higher frequencies (22, 43, and 86\,GHz) show a behavior consistent with a circular structure of the magnetic field.
To our knowledge, this is the first time that a jet source observed within a very small angle to the line of sight shows direct evidence of the toroidal component of the magnetic field in an extragalactic relativistic jet.

\subsubsection{Maps of linear polarization degree}
\begin{figure}
   \centering
       \includegraphics[height=18.cm]{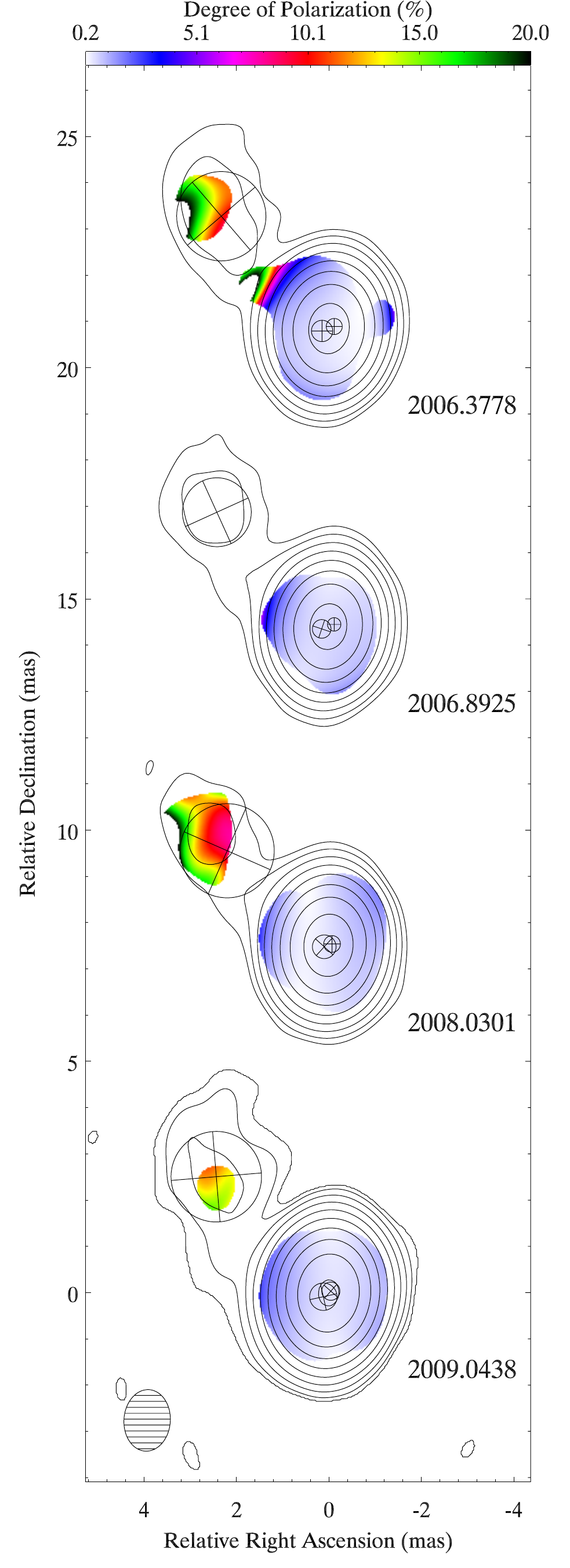}
       \caption{Linear polarization degree of NRAO\,150 as measured with the VLBA at 8\,GHz. Contour levels represent the total intensity distribution of the source for every observing epoch as shown in Fig.~\ref{maps8}.}
       \label{pol_8}
   \end{figure}

\begin{figure}
   \centering
       \includegraphics[height=21.cm]{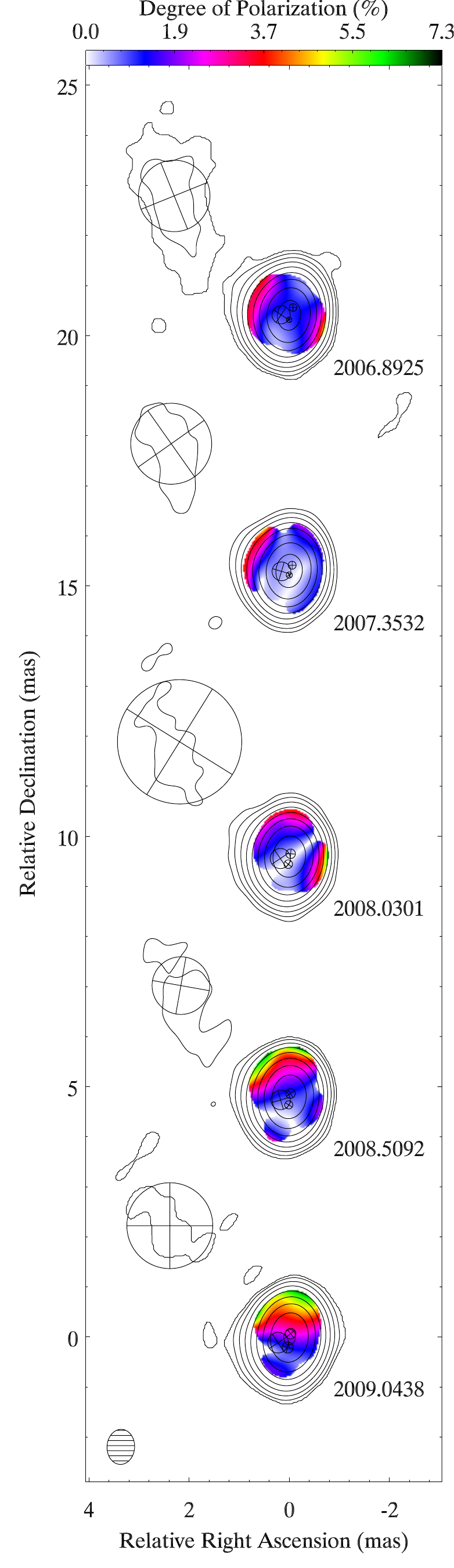}
       \caption{Same as Fig.~\ref{pol_8}, but for 15\,GHz observations.}
       \label{pol_15}
   \end{figure}

\begin{figure}
   \centering
       \includegraphics[height=21.cm]{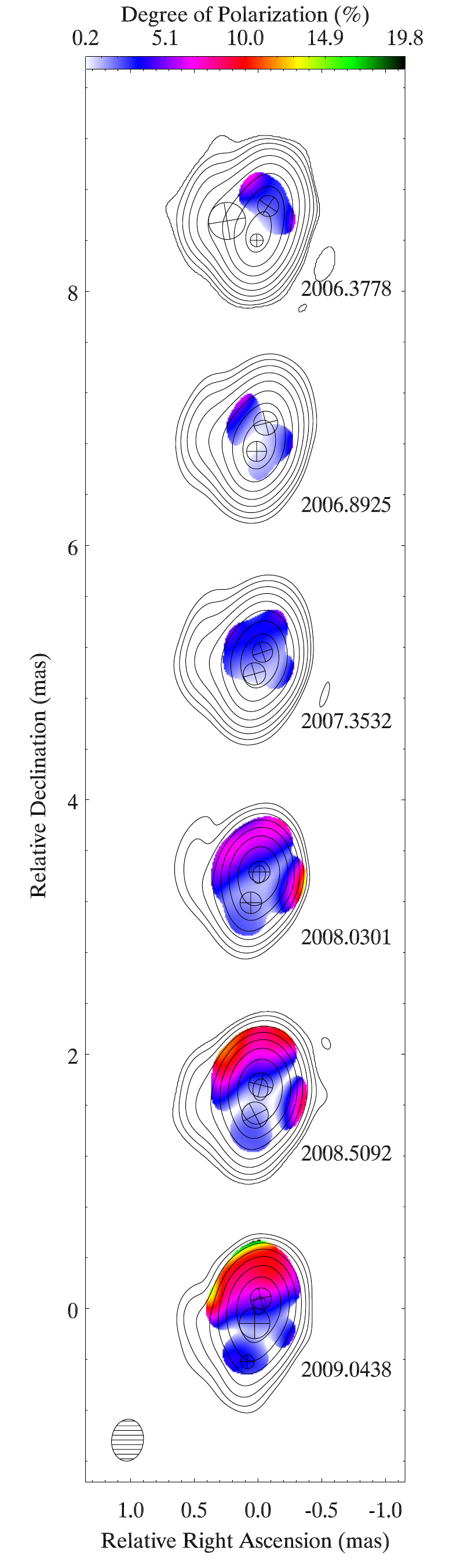}
       \caption{Same as Fig.~\ref{pol_8}, but for 22\,GHz observations.}
       \label{pol_22}
   \end{figure}

\begin{figure}
   \centering
       \includegraphics[height=21.cm]{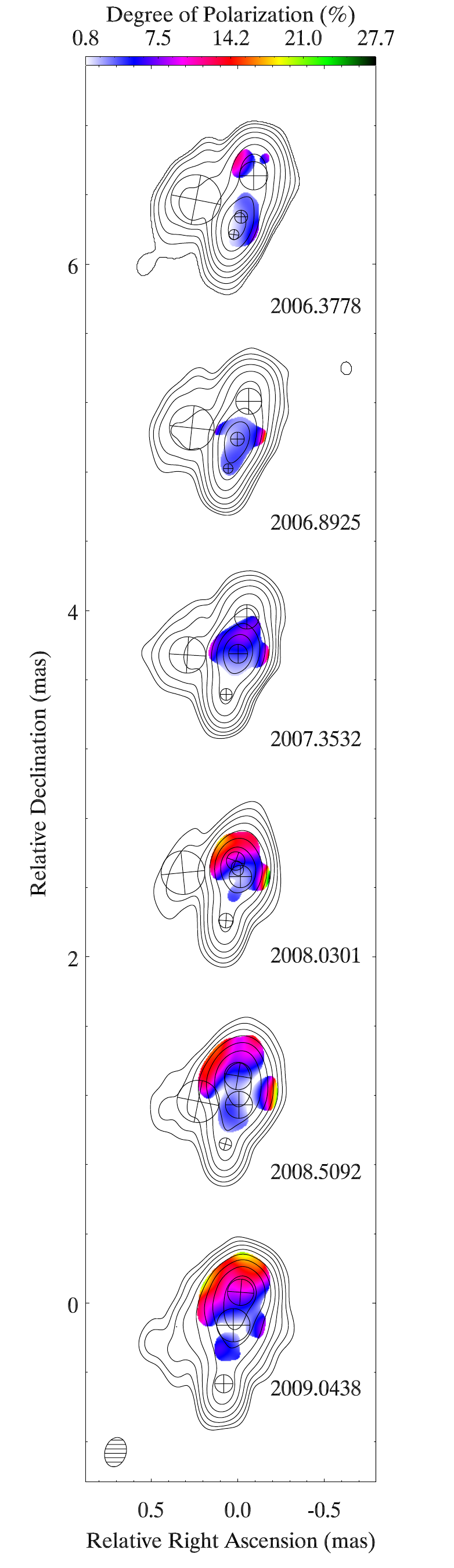}
       \caption{Same as Fig.~\ref{pol_8}, but for 43\,GHz observations.}
       \label{pol_43}
   \end{figure}

Figures~\ref{pol_8} to \ref{pol_43} show maps of linear polarization degree for every one of our observing frequencies from 8.4\,GHz to 43\,GHz. 
We do not include 86\,GHz maps of the polarization degree because there is no easy way to obtain a reliable result. 
The reason for this is that the 86\,GHz polarized intensity distribution is slightly shifted with respect to the total intensity, which results in degrees of polarization maps with fast increments towards the edge of the jet. If we flag these regions, we lose almost all polarization degree emission. For this reason, we conclude that the polarization-degree maps at this frequency do not show confident results. 
A possible cause for these shifts in the polarization structure may be introduced by inaccuracies in the determination of the 86\,GHz D-terms.


The 8\,GHz maps of linear polarization degree, $p$, (Fig.~\ref{pol_8}) show that polarization is very low and homogeneous in the central region near Q0 with maximum polarization degrees $p\lesssim2\%$.
In optically thin regions we found maximum values of roughly 15 $\%$, but the total and polarized flux are very low in these regions so these values could be in the limit of noise showing higher values than the real ones. 

Together with the evident progressive $p$ increase in the region surrounding Q0 from 2006 to 2009, clearly observed at observing frequencies higher than 8\,GHz, we also report an increase of linear polarization degree with observing frequency. 
The maximum $p$ values during epoch 2009.04 in the northern region are $\sim5\%$ at 15\,GHz, $\sim10\%$ at 22\,GHz, and $\sim15\%$ at 43\,GHz.

It is well established that the degree of polarization depends on the optical depth \citep{Pacholczyk1970} and increases in optically thin regions. 
However, the spectral index maps in 2009.4 and adjacent epochs do not show clear evidence of the significantly more optically thin nature of the source regions close to Q0, e.g., compared to those around Q3 (or Q2 in the case of the 15\,GHz to 22\,GHz spectral index map).
Furthermore, Fig.~\ref{sp_epocaF} shows that the peak of NRAO\,150's spectrum in 2009 is located at $\sim15$\,GHz, because the integrated emission at 22\,GHz and 43\,GHz is optically thin. 
Therefore, synchrotron self-absorption alone is not sufficient to explain the increase of degree of polarization with frequency and the very low degree of polarization in optically thin regions.

A relevant component of the actual magnetic field of the jet plasma with a tangled orientation can produce depolarization in optically thin regions.
Moreover, beam depolarization is an additional effect that may play a relevant role.
The combination of a toroidal magnetic field configuration into a projected circular structure of $\sim0.3$\,mas diameter, as suggested by the observations reported in the previous section, together with the use of different observing beams because the low frequency ones are considerably larger than this circular structure, certainly produces more efficient depolarization for larger beams (i.e., larger observing wavelengths).
This is even true for VLBI observations at shorter wavelengths if the observing beam is not negligible compared to the size of the coherent polarization structure of the source. This seems to be the case at least for our 22\,GHz polarimetric observations.
Beam depolarization can also explain the very low linear polarization degree ($<$ 2 $\%$) observed at 8\,GHz.

\subsection{Kinematic analysis}
To increase the time span in our study of the kinematical behavior in NRAO 150 we use the data from the 34 VLBA images at 43\,GHz presented by \cite{Agudo07}.

Following \cite{Agudo07}, we have tested the trajectories of the tracked model-fit components by assuming Q0 to be the reference center of the kinematic system, i.e., the position of Q0 is set to the (0,0) position for every observing epoch.  
Here we add the position of components fitted in the new images presented in this work, which --under the above mentioned assumption for Q0-- gives the kinematical behavior shown in Fig.~\ref{fit}-a. 
The symbols (crosses, squares, and asterisks) represent the position of components with respect to the Q0 feature. 
For Fig.~\ref{fit}, we did not use the data corresponding to the observing epochs in 2008, since the source region around Q0 is strongly influenced by the presence of the Qn component, and therefore the position of Q0 is not reliable enough for a kinematical study in such epochs.
Figure~\ref{fit}-a makes evident the wobbling on the plane of the sky of the emission features in NRAO\,150 which continues its counterclockwise rotation, as reported by \cite{Agudo07}, and without any sign of changing the sense of rotation. 
This implies that if there is any periodicity in the behavior of the source (which cannot be assessed by the data we have compiled so far), it cannot have a period smaller than around 12 years --the maximum time span covered by this study and the \cite{Agudo07} study together. 

When compared to the remaining emission components in NRAO\,150, Qn is a rather peculiar emission feature. 
First of all, Qn has a drastically different speed, with a mean proper motion --measured considering Q0 as reference -- of 0.09$\pm$0.01 mas/yr (6.3$\pm$1.1 c), while the velocities measured for Q1, Q2, and Q3 are 3.26$\pm$0.14 c, 2.85$\pm$0.07 c, and 2.29$\pm$0.14 c, respectively \citep{Agudo07}. 
In addition, as reported in Section 3.4, the EVPA in the region surrounding Qn is always parallel to its direction of propagation and the polarization degree increases as Qn moves closer to Q0. 
Therefore, the properties of Qn suggest that this emission feature is related to a jet perturbation, whose nature is drastically different from that of the remaining components in the jet.
That is, it propagates with a range of speeds much wider than other jet features; it is the only jet component that moves with a south to north trajectory, and it shows much stronger polarization intensity (and perhaps magnetic field order) than the remaining polarization regions in NRAO\,150.

\section{A new--alternative kinematic scenario: Jet internal rotation}

All evidence reported so far regarding the properties of NRAO\,150 indicate that the jet points at an extremely small angle to the line of sight. 
Our spectral index study does not allow us to unambiguously identify a stationary core from where to reference a kinematic study of the jet in NRAO\,150.
This may be consistent with a scenario where all observed emission features at frequencies higher than 22\,GHz move with prominent emission eclipsing (therefore avoiding the detection of) the actual core on inner jet regions. 
Moreover, we have shown that at epochs when the linear polarization degree is large ($\gtrsim10^{\circ}$) and therefore indicative of a better ordered magnetic field, our observations suggest that we observe the toroidal structure of the magnetic field threading the jet plasma rotating around its axis.
Finally, we observe a rather low degree of polarization that increases with observing frequency. 
All this evidence can be explained together by a scenario where a toroidal magnetic field configuration in the jet is observed under a very small angle from the line of sight.

Moreover, we have measured the rotation of the total--flux emission structure of NRAO\,150 between epochs in a completely independent way compared to previous studies, and without making any assumption about the actual position of either the core or any other emission feature. 
In this section we present an alternative kinematic scenario to explain the behavior of NRAO\,150, as seen at high frequencies with VLBI, by taking into account the above mentioned evidence.  
In particular, unlike previous studies of NRAO\,150, none of the positions of the fitted emission components will be assumed to remain stationary in the jet.

We assume that the innermost jet emission regions move rotating around the jet axis when the jet is seen face on --which is approximately the case of NRAO\,150, see above.
These trajectories may be produced by a helical or quasi-helical magnetic field threading the innermost, magnetically dominated regions of the jet. 
If this is the case, the material has to follow the field lines, hence also tracing bent trajectories around the jet axis. 
If the jet is seen face-on during the evolution of the main emission features traveling outwards from the innermost regions, these features should be observed rotating around a fixed point, i.e., around the actual jet axis as seen in projection on the plane of the sky.

\begin{figure}
   \centering
       \includegraphics[width=9cm]{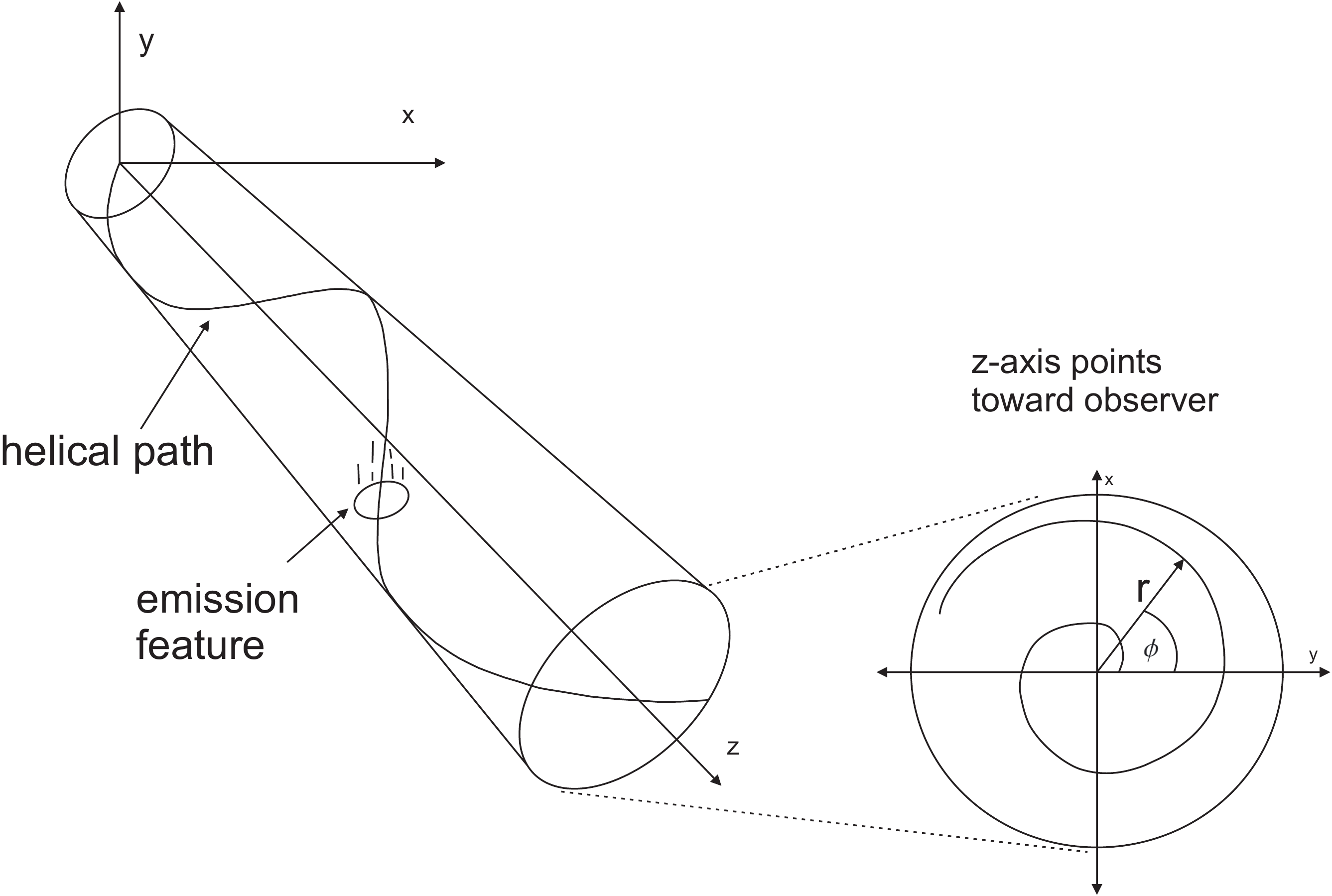}
       \caption{Conceptual representation of the new model proposed to explain the bent trajectories of emission features in the 43\,GHz images of NRAO\,150. The plot to the right represents  the trajectory of an emission feature when the z-axis points towards the observer within a very small angle from the line of sight.}
       \label{esqueme}
   \end{figure}

Figure~\ref{esqueme} shows a conceptual scheme of this kinematic scenario, in which the z-axis points towards the observer within a very small (assumed negligible) angle from the line of sight. 
We describe this kinematic scenario by making use of the expressions in polar coordinates,
   \begin{equation}
    \centering
     r_{i}(t)= r_{i}^{ini} + v_{i}^{r} \hspace{0.15cm} t   
     \end{equation} 
     \begin{equation}
     \centering 
      \phi_{i}(t)=\omega_{i} \hspace{0.1cm} t  +  \phi_{i}^{ini},
      \end{equation}   
where $r_{i}(t)$ is proportional to the radial velocity $v_{i}^{r}$ (that we assume constant, but different for each component), $r_{i}^{ini}$ is the distance from the jet axis at time $t=0$, and $\phi_{i}(t)$ is the angle measured in the x$-$y plane starting from $\phi_{i}^{ini}$ at $t=0$. This angle varies in time, depending on the angular velocity, $\omega_{i}$, which is also assumed constant, but different for every emission feature. Subscript i refers to each emission feature.

\begin{figure}
   \centering
       \includegraphics[width=9cm]{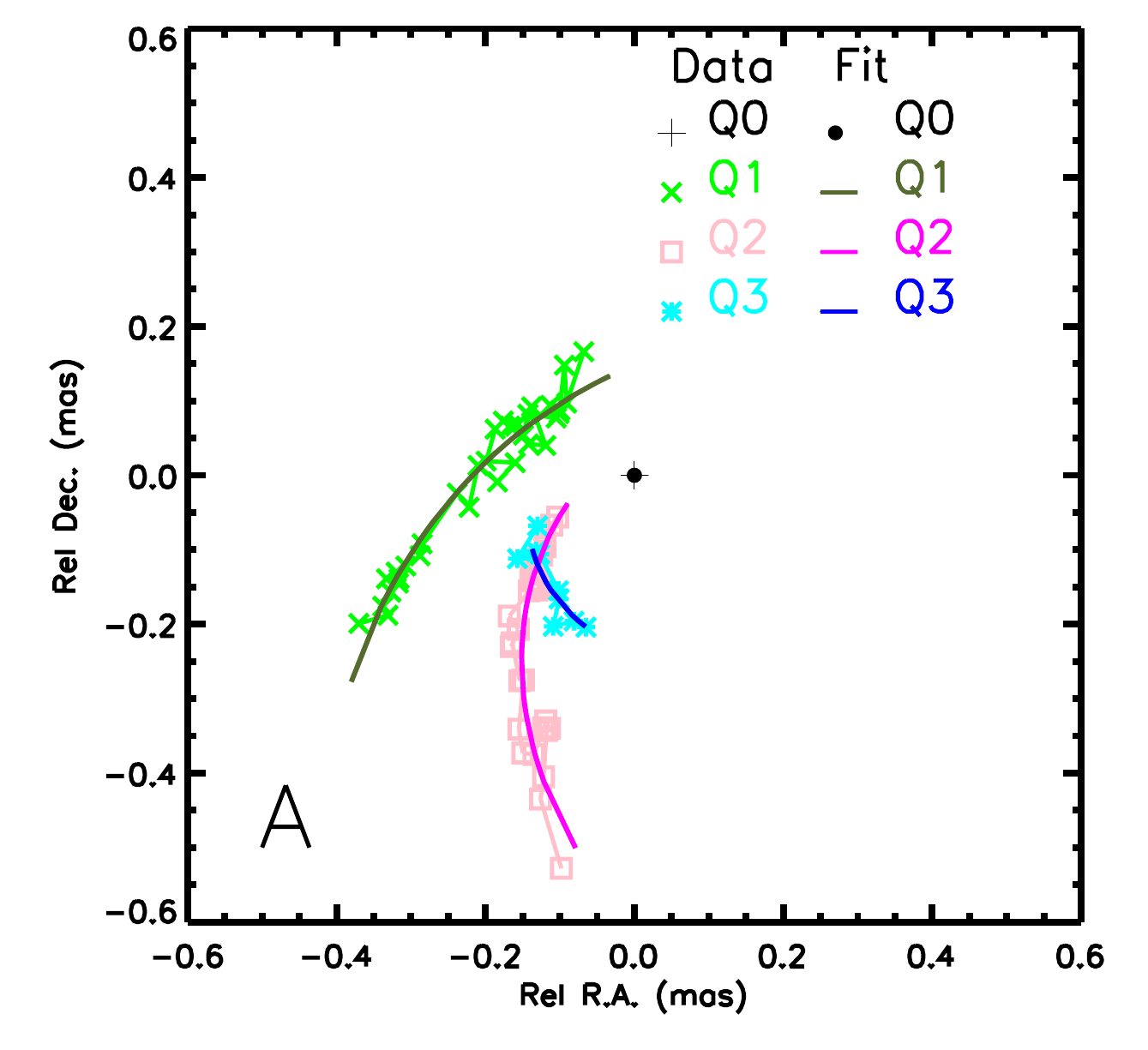}
       \includegraphics[width=9cm]{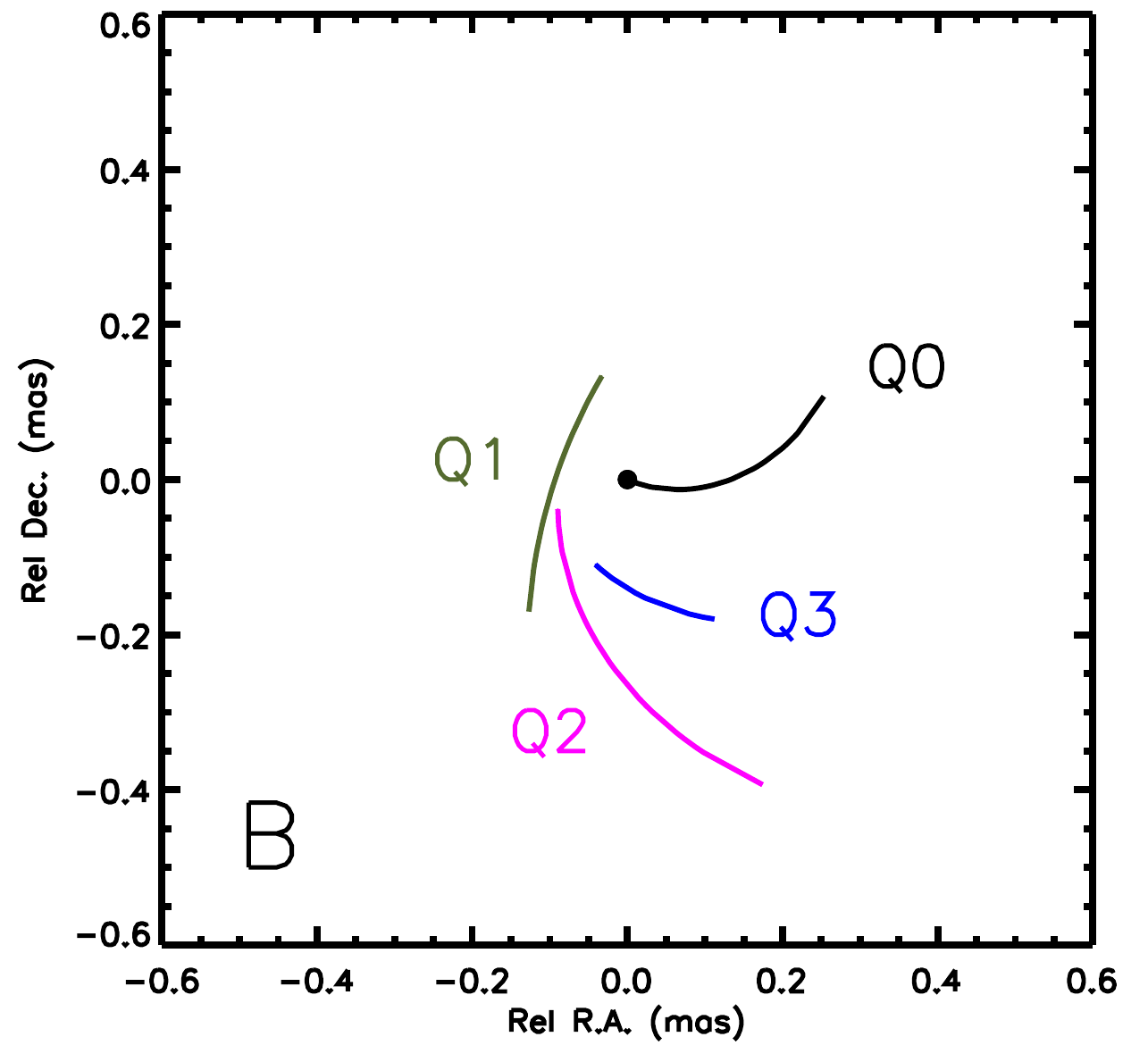}
       \caption{ \emph{a)} Positions of 43\,GHz model fitted components as observed on the plane of the sky when the Q0 component is considered to remain stationary at (0,0). Green crosses, pink squares, and cyan asterisks represent the positions of Q1, Q2, and Q3 components, respectively. Curved lines represent the best-fit trajectories of emission features discussed in the text.  
                     \emph{b)} Best-fit trajectories of emission features as given by our new kinematic scenario discussed in the text. The black line symbolizes the trajectory of Q0 (now describing a bent trajectory around the new (0,0) position), whereas the dark green, magenta, and blue lines represent the trajectories of Q1, Q2, and Q3, respectively.}
        \label{fit}
   \end{figure}

  Given that the position of our observed components all refer to the position of component Q0, which for simplicity was assumed to remain stationary (see Fig.~\ref{fit}-a), in order to compare with observations we have first to refer the positions in our model to that of Q0, so that in Cartesian coordinates we have

  \begin{equation}
     x_{i}(t)=r_{i}(t)\hspace{0.1cm} \cos( \phi_{i}(t) ) - r_{0}(t)\hspace{0.1cm} \cos( \phi_{0}(t) ) \\     
   \end{equation}
  \begin{equation}
     y_{i}(t)=r_{i}(t)\hspace{0.1cm} \sin(  \phi_{i}(t)  ) - r_{0}(t)\hspace{0.1cm} \sin( \phi_{0}(t)  ), \\ 
   \end{equation}  
where the subscript 0 refers to Q0 parameters and subscript i refers to the rest of the components (e.g., Q1, Q2, Q3) .

We used a $\chi^2$ minimization scheme to look for the best-fit values of $r_{i}^{ini}$, $v_{i}^{r}$, $ \phi_{i}^{ini}$, and $\omega_{i}$ for the trajectories of every one of the emission features under study.
All emission features seen at 43\,GHz (i.e., Q0, Q1, Q2, and Q3) were fitted simultaneously.
The range of explored values for the kinematic fitting was sufficiently broad to be sure that we include all possible kinematic behaviors compatible with the observations. 
In particular, $r_{i}^{ini}$ was allowed to take values from 0.0\,mas to 0.9\,mas with increments of 0.01, $v_{i}^{r}$ ranged from 0.0\,mas/yr to 0.1\,mas/yr with increments of 0.0025, $\omega_{i}$ from 0.0$^{\circ}$/yr to 19.4$^{\circ}$/yr with increments of 0.48 (for all fit components, except Q1 which uses 0.28), and $\phi_{i}^{ini}$ from 0.57$^{\circ}$ to 360$^{\circ}$ with increments of 5.7 (for all fit components, except Q1 which uses 11.45).
We note, as for Q1, Q2, and Q3, all fitting parameters were let free and fitted to the best values reproducing the trajectory of Q0, therefore allowing this emission feature to move on the plane of the sky with regard to an a priori unknown kinematic center.

The trajectories provided by our best fit, with a reduced chi square of 5.46, are graphically represented for Q0, Q1, Q2, and Q3 in Fig.~\ref{fit}-b, whereas the corresponding fitting parameters are shown in Table~\ref{Table_1}. 
Figure~\ref{fit} shows that our method is able to accurately reproduce the trajectory of the most prominent emission features in our 43\,GHz images, which supports the idea that our model can explain the most salient properties of the kinematical behavior of NRAO\,150.
This also includes Q0, for which a fast rotational motion around the new kinematic center, and in the same counterclockwise sense as for the Q2 and Q3 emission features, is suggested by our fit.
In contrast, the fitted trajectory of Q1 is almost rectilinear. 
Table~\ref{Table_1} shows a very small angular speed for Q1, whereas the remaining emission features rotate around the (0,0) position in Fig.~\ref{fit}-b with considerably larger angular speeds.

We estimated the apparent speeds of the 43\,GHz emission features, under the new kinematic scenario suggested by our fit, in the same way as in \cite{Homan01},  \cite{Jorstad05}, and \cite{Agudo07}.
This is, by fitting the trajectory of every emission feature, as given by our rotation model, through a polynomial.
We employed a second-order polynomial for all emission features in this work.
The mean measured proper motions are 0.0253$\pm$0.0015 mas/yr, 0.0276$\pm$0.0019 mas/yr, 0.039$\pm$0.00069 mas/yr, and 0.05$\pm$0.004 mas/yr for Q0, Q1, Q2, and Q3, respectively. 
These values correspond to superluminal apparent speeds of 1.77$\pm$0.10 c, 1.93$\pm$0.13 c, 2.73$\pm$0.048 c, and 3.5$\pm$0.28 c.

By decomposing the mean projected speed into their radial and non-radial directions we obtain non-radial speeds of 1.60$\pm$0.16 c, 0.19$\pm$3.96 c, 1.55$\pm$0.32 c, and 2.93$\pm$1.47 c for Q0, Q1, Q2, and Q3, respectively. 
Therefore, as under the assumptions for the stationary position of Q0 made in \cite{Agudo07}, our new kinematic model yields superluminal apparent velocities in the non-radial direction of propagation of emission features. 
This points out the remarkable non-ballistic properties of the innermost emission regions in NRAO\,150, as well as the presence of strong, non-rectilinear, magnetic fields in the relevant emission regions.

We stress that the new kinematic scenario outlined above is based on the observational evidence presented in this paper for the first time.
In addition, our best fit to such a simple kinematic-scenario describes rather accurately the observed trajectories of the most prominent 43\,GHz model-fit components in NRAO\,150, which further supports the initial hypothesis that we are observing the actual rotation of the emitting plasma around its jet axis from a very small (almost negligible) angle of the line of sight to this jet axis.
For that, a strong toroidal magnetic field component in the relevant emission regions is required, which is fully consistent with both the high-frequency EVPA distribution observed from 22\,GHz to 86\,GHz (see Section 3.4.1) and with a helical structure of the magnetic field as suggested before for other sources both from direct observations \citep{Zamanina13} or indirect methods \citep{Asada02, Gabuzda05, Hardee05, Gomez08}.


\begin{table}
\centering
\caption{Best-fit parameters corresponding to the new kinematic model for the 43\,GHz structure of NRAO\,150 discussed in the text.} 
\label{Table_1}   

 \resizebox*{0.5\textwidth}{!}{
\begin{tabular}{ccccc}
\hline
Comp & $r^{ini}$ &   $v^{r} $         &    $\phi^{ini}$     &  $\omega$    \\
          &  [mas]  &   [mas$/$yr]  &   [$^o$]       &   [$^o/$yr]     \\\hline

Q0  &  0.16  $\pm $  0.01 &  0.010  $\pm $ 0.002  & 276.1   $\pm $  5.7      & 6.33   $\pm $  0.48 \\
Q1  & 0.03  $\pm $   0.02 &  0.027  $\pm $ 0.004  &  238.3  $\pm $  11.4    & 1.14   $\pm $  1.08 \\
Q2  &  0.21  $\pm $  0.01&   0.032  $\pm $ 0.002   & 249.8  $\pm $  5.7      & 3.40  $\pm $  0.48 \\
Q3  &  0.15  $\pm $  0.03 &  0.027  $\pm $  0.005  & 231.4  $\pm $  6.8      & 7.67 $\pm $ 1.03 \\\hline
\end{tabular}
}
\end{table}

\section{Summary}

We have employed multi-frequency polarimetric VLBI observations at 8, 15, 22, 43, and 86\,GHz to study the jet emitting region of NRAO\,150, with special emphasis on the innermost emitting regions detected at 22, 43, and 86\,GHz.
Our spectral index study suggests that there is not a preferential optically thick region that could be assumed as the millimeter core of the jet. We note, however, that the spectral index maps at the higher frequencies (43\,GHz and 86\,GHz) may be significantly affected by the uncertainties in the image alignment.
Our cross-correlation method confirms that the 43\,GHz emitting regions in NRAO\,150 rotate at high speeds on the plane of the sky with respect to a reference point that does not need to be related to any modeled jet feature.
We measure a rather low linear polarization degree, even for optically thin regions, at all observing frequencies. 
This is consistent with the hypothesis that we are seeing the jet in NRAO\,150 oriented almost face on, therefore enhancing the depolarization effect as integrated over the line of sight.
The degree of polarization also increases with observing frequency in optically thin jet regions. 
The observed EVPA distribution at frequencies higher than 22\,GHz during observing epochs --where the higher linear polarization degrees are detected-- suggests that we have detected for the first time the toroidal component of the magnetic field threading the jet plasma on jet regions where the jet cross section is on the order of $\sim0.3$\,mas (which correspond to $\sim2.55$\,pc).
The observed toroidal polarization structure is also consistent with the increasing polarization degree for increasing angular resolution (and observing frequency).
This produces a higher degree of polarization at progressively smaller scales, whereas the integrated polarization intensity produced by the toroidal field is cancelled out at lower angular resolution when averaged inside the larger observing beam at longer observing wavelengths.

All these observational lines of evidence, together with the previously reported extreme jet-wobbling in NRAO\,150, are fully consistent with a new kinematic scenario where none of the emission features reported at 43\,GHz need to be considered the kinematic center from where the jet cross section rotates.
Within this scenario, the main kinematic and polarization properties of the 43\,GHz emitting structure of NRAO\,150 are explained by the internal rotation of such emission regions around the jet axis when the jet is seen almost face on.
A simplified model developed to fit helical trajectories to the observed kinematics of the 43\,GHz features under this scenario fully supports this hypothesis.
This not only explains the kinematics of the innermost regions of the jet in NRAO\,150 in terms of internal jet rotation, but it also opens a new possibility to interpret the observed jet wobbling in the innermost regions of other blazars.

\begin{acknowledgements}
We would like to thank the referee for his/her comments that have helped to improve the paper.
This research has been supported by the Spanish Ministry of Science and Innovation grants AYA2010-14844, by the Regional Government of Andaluc\'ia (Spain) grant P09-FQM-4784. 
This paper is partially based on observations carried out with the VLBA, the MPIfR 100 m Effelsberg Radio Telescope, the IRAM Plateau de Bure Millimetre Interferometer, the IRAM 30 m Millimetre Telescope, the Onsala 20 m Radio Telescope, and the Mets\"ahovi 14 m Radio Telescope. 
This research has also made use of data from VLA and Mojave (Monitoring Of Jets in Active galactic nuclei with VLBA Experiments) Program.
The VLBA and VLA are instruments of the National Radio Astronomy Observatory, a facility of the National Science Foundation of the USA operated under cooperative agreement by Associated Universities, Inc. (USA). IRAM is supported by MPG (Germany), INSU/CNRS (France), and IGN (Spain). The GMVA is operated by the MPIfR, IRAM, NRAO, OSO, and MRO. We thank the staff of the participating observatories for their efficient and continuous support. The MOJAVE project was supported under NASA-Fermi grants NNX08AV67G and 11-Fermi11-0019.

\end{acknowledgements}

\bibliographystyle{aa}

\bibliography{S.N.Molina}

\begin{thebibliography}{39}
\expandafter\ifx\csname natexlab\endcsname\relax\def\natexlab#1{#1}\fi

\bibitem[{{Acosta-Pulido} {et~al.}(2010){Acosta-Pulido}, {Agudo}, {Barrena},
  {Ramos Almeida}, {Manchado}, \& {Rodr{\'{\i}}guez-Gil}}]{Acosta10}
{Acosta-Pulido}, J.~A., {Agudo}, I., {Barrena}, R., {et~al.} 2010, \aap, 519,
  A5

\bibitem[{{Agudo} {et~al.}(2007){Agudo}, {Bach}, {Krichbaum}, {Marscher},
  {Gonidakis}, {Diamond}, {Perucho}, {Alef}, {Graham}, {Witzel}, {Zensus},
  {Bremer}, {Acosta-Pulido}, \& {Barrena}}]{Agudo07}
{Agudo}, I., {Bach}, U., {Krichbaum}, T.~P., {et~al.} 2007, \aap, 476, L17

\bibitem[{{Agudo} {et~al.}(2006){Agudo}, {G{\'o}mez}, {Gabuzda}, {Marscher},
  {Jorstad}, \& {Alberdi}}]{Agudo06}
{Agudo}, I., {G{\'o}mez}, J.~L., {Gabuzda}, D.~C., {et~al.} 2006, \aap, 453,
  477

\bibitem[{{Agudo} {et~al.}(2012){Agudo}, {Marscher}, {Jorstad}, {G{\'o}mez},
  {Perucho}, {Piner}, {Rioja}, \& {Dodson}}]{Agudo12}
{Agudo}, I., {Marscher}, A.~P., {Jorstad}, S.~G., {et~al.} 2012, \apj, 747, 63

\bibitem[{{Asada} {et~al.}(2002){Asada}, {Inoue}, {Uchida}, {Kameno},
  {Fujisawa}, {Iguchi}, \& {Mutoh}}]{Asada02}
{Asada}, K., {Inoue}, M., {Uchida}, Y., {et~al.} 2002, \pasj, 54, L39

\bibitem[{{Beckert} \& {Falcke}(2002)}]{Beckert02}
{Beckert}, T. \& {Falcke}, H. 2002, \aap, 388, 1106

\bibitem[{{Croke} \& {Gabuzda}(2008)}]{Croke08}
{Croke}, S.~M. \& {Gabuzda}, D.~C. 2008, \mnras, 386, 619

\bibitem[{{Croke} {et~al.}(2010){Croke}, {O'Sullivan}, \& {Gabuzda}}]{Croke10}
{Croke}, S.~M., {O'Sullivan}, S.~P., \& {Gabuzda}, D.~C. 2010, \mnras, 402, 259

\bibitem[{{Gabuzda}(2005)}]{Gabuzda05}
{Gabuzda}, D.~C. 2005, in Astronomical Society of the Pacific Conference
  Series, Vol. 345, From Clark Lake to the Long Wavelength Array: Bill
  Erickson's Radio Science, ed. N.~{Kassim}, M.~{Perez}, W.~{Junor}, \&
  P.~{Henning}, 264

\bibitem[{{G{\'o}mez} {et~al.}(2002){G{\'o}mez}, {Marscher}, {Alberdi},
  {Jorstad}, \& {Agudo}}]{Gomez02_b}
{G{\'o}mez}, J.-L., {Marscher}, A.~P., {Alberdi}, A., {Jorstad}, S.~G., \&
  {Agudo}, I. 2002, in , VLBA Scientific Memo, 30.

\bibitem[{{G{\'o}mez} {et~al.}(2008){G{\'o}mez}, {Marscher}, {Jorstad},
  {Agudo}, \& {Roca-Sogorb}}]{Gomez08}
{G{\'o}mez}, J.~L., {Marscher}, A.~P., {Jorstad}, S.~G., {Agudo}, I., \&
  {Roca-Sogorb}, M. 2008, \apjl, 681, L69

\bibitem[{{G{\'o}mez} {et~al.}(2011){G{\'o}mez}, {Roca-Sogorb}, {Agudo},
  {Marscher}, \& {Jorstad}}]{Gomez11}
{G{\'o}mez}, J.~L., {Roca-Sogorb}, M., {Agudo}, I., {Marscher}, A.~P., \&
  {Jorstad}, S.~G. 2011, \apj, 733, 11

\bibitem[{{Hardee} {et~al.}(2005){Hardee}, {Walker}, \& {G{\'o}mez}}]{Hardee05}
{Hardee}, P.~E., {Walker}, R.~C., \& {G{\'o}mez}, J.~L. 2005, \apj, 620, 646

\bibitem[{{Homan} {et~al.}(2001){Homan}, {Ojha}, {Wardle}, {Roberts}, {Aller},
  {Aller}, \& {Hughes}}]{Homan01}
{Homan}, D.~C., {Ojha}, R., {Wardle}, J.~F.~C., {et~al.} 2001, \apj, 549, 840

\bibitem[{{Jorstad} {et~al.}(2005){Jorstad}, {Marscher}, {Lister}, {Stirling},
  {Cawthorne}, {Gear}, {G{\'o}mez}, {Stevens}, {Smith}, {Forster}, \&
  {Robson}}]{Jorstad05}
{Jorstad}, S.~G., {Marscher}, A.~P., {Lister}, M.~L., {et~al.} 2005, \aj, 130,
  1418

\bibitem[{{Komatsu} {et~al.}(2009){Komatsu}, {Dunkley}, {Nolta}, {Bennett},
  {Gold}, {Hinshaw}, {Jarosik}, {Larson}, {Limon}, {Page}, {Spergel},
  {Halpern}, {Hill}, {Kogut}, {Meyer}, {Tucker}, {Weiland}, {Wollack}, \&
  {Wright}}]{Komatsu09}
{Komatsu}, E., {Dunkley}, J., {Nolta}, M.~R., {et~al.} 2009, \apjs, 180, 330

\bibitem[{{Leppanen} {et~al.}(1995){Leppanen}, {Zensus}, \&
  {Diamond}}]{Leppanen95}
{Leppanen}, K.~J., {Zensus}, J.~A., \& {Diamond}, P.~J. 1995, \aj, 110, 2479

\bibitem[{{Lister} {et~al.}(2013){Lister}, {Aller}, {Aller}, {Homan},
  {Kellermann}, {Kovalev}, {Pushkarev}, {Richards}, {Ros}, \&
  {Savolainen}}]{Lister13}
{Lister}, M.~L., {Aller}, M.~F., {Aller}, H.~D., {et~al.} 2013, \aj, 146, 120

\bibitem[{{Lister} {et~al.}(2009){Lister}, {Cohen}, {Homan}, {Kadler},
  {Kellermann}, {Kovalev}, {Ros}, {Savolainen}, \& {Zensus}}]{Lister09}
{Lister}, M.~L., {Cohen}, M.~H., {Homan}, D.~C., {et~al.} 2009, \aj, 138, 1874

\bibitem[{{Lister} {et~al.}(2003){Lister}, {Kellermann}, {Vermeulen}, {Cohen},
  {Zensus}, \& {Ros}}]{Lister03}
{Lister}, M.~L., {Kellermann}, K.~I., {Vermeulen}, R.~C., {et~al.} 2003, \apj,
  584, 135

\bibitem[{{Lobanov} \& {Roland}(2005)}]{Lobanov05}
{Lobanov}, A.~P. \& {Roland}, J. 2005, \aap, 431, 831

\bibitem[{{Marscher} {et~al.}(2008){Marscher}, {Jorstad}, {D'Arcangelo},
  {Smith}, {Williams}, {Larionov}, {Oh}, {Olmstead}, {Aller}, {Aller},
  {McHardy}, {L{\"a}hteenm{\"a}ki}, {Tornikoski}, {Valtaoja}, {Hagen-Thorn},
  {Kopatskaya}, {Gear}, {Tosti}, {Kurtanidze}, {Nikolashvili}, {Sigua},
  {Miller}, \& {Ryle}}]{Marscher08_Nat}
{Marscher}, A.~P., {Jorstad}, S.~G., {D'Arcangelo}, F.~D., {et~al.} 2008, \nat,
  452, 966

\bibitem[{{Mart{\'{\i}}-Vidal} {et~al.}(2012){Mart{\'{\i}}-Vidal}, {Krichbaum},
  {Marscher}, {Alef}, {Bertarini}, {Bach}, {Schinzel}, {Rottmann}, {Anderson},
  {Zensus}, {Bremer}, {Sanchez}, {Lindqvist}, \& {Mujunen}}]{Marti12}
{Mart{\'{\i}}-Vidal}, I., {Krichbaum}, T.~P., {Marscher}, A., {et~al.} 2012,
  \aap, 542, A107

\bibitem[{{Mart{\'{\i}}-Vidal} {et~al.}(2011){Mart{\'{\i}}-Vidal}, {Marcaide},
  {Alberdi}, {P{\'e}rez-Torres}, {Ros}, \& {Guirado}}]{Marti11}
{Mart{\'{\i}}-Vidal}, I., {Marcaide}, J.~M., {Alberdi}, A., {et~al.} 2011,
  \aap, 533, A111

\bibitem[{{McKinney} \& {Blandford}(2009)}]{McKinney09}
{McKinney}, J.~C. \& {Blandford}, R.~D. 2009, \mnras, 394, L126

\bibitem[{{Meier} {et~al.}(2001){Meier}, {Koide}, \&
  {Uchida}}]{Meier01_Science}
{Meier}, D.~L., {Koide}, S., \& {Uchida}, Y. 2001, Science, 291, 84

\bibitem[{{Mizuno} {et~al.}(2012){Mizuno}, {Lyubarsky}, {Nishikawa}, \&
  {Hardee}}]{Mizuno12}
{Mizuno}, Y., {Lyubarsky}, Y., {Nishikawa}, K.-I., \& {Hardee}, P.~E. 2012,
  \apj, 757, 16

\bibitem[{{Mutel} \& {Denn}(2005)}]{Mutel05}
{Mutel}, R.~L. \& {Denn}, G.~R. 2005, \apj, 623, 79

\bibitem[{{O'Sullivan} \& {Gabuzda}(2009)}]{OSullivan09}
{O'Sullivan}, S.~P. \& {Gabuzda}, D.~C. 2009, \mnras, 393, 429

\bibitem[{{Pacholczyk}(1970)}]{Pacholczyk1970}
{Pacholczyk}, A.~G. 1970, {Radio astrophysics. Nonthermal processes in galactic
  and extragalactic sources}

\bibitem[{{Perucho} {et~al.}(2012){Perucho}, {Kovalev}, {Lobanov}, {Hardee}, \&
  {Agudo}}]{Perucho12}
{Perucho}, M., {Kovalev}, Y.~Y., {Lobanov}, A.~P., {Hardee}, P.~E., \& {Agudo},
  I. 2012, \apj, 749, 55

\bibitem[{{Savolainen} {et~al.}(2006){Savolainen}, {Wiik}, {Valtaoja},
  {Kadler}, {Ros}, {Tornikoski}, {Aller}, \& {Aller}}]{Savolainen06}
{Savolainen}, T., {Wiik}, K., {Valtaoja}, E., {et~al.} 2006, \apj, 647, 172

\bibitem[{{Shepherd}(1997)}]{Shepherd97}
{Shepherd}, M.~C. 1997, in Astronomical Society of the Pacific Conference
  Series, Vol. 125, Astronomical Data Analysis Software and Systems VI, ed.
  G.~{Hunt} \& H.~{Payne}, 77

\bibitem[{{Steffen} {et~al.}(1995){Steffen}, {Zensus}, {Krichbaum}, {Witzel},
  \& {Qian}}]{Steffen95}
{Steffen}, W., {Zensus}, J.~A., {Krichbaum}, T.~P., {Witzel}, A., \& {Qian},
  S.~J. 1995, \aap, 302, 335

\bibitem[{{Stirling} {et~al.}(2003){Stirling}, {Cawthorne}, {Stevens},
  {Jorstad}, {Marscher}, {Lister}, {G{\'o}mez}, {Smith}, {Agudo}, {Gabuzda},
  {Robson}, \& {Gear}}]{Stirling03}
{Stirling}, A.~M., {Cawthorne}, T.~V., {Stevens}, J.~A., {et~al.} 2003, \mnras,
  341, 405

\bibitem[{{Thum} {et~al.}(2008){Thum}, {Wiesemeyer}, {Paubert}, {Navarro}, \&
  {Morris}}]{Thum08}
{Thum}, C., {Wiesemeyer}, H., {Paubert}, G., {Navarro}, S., \& {Morris}, D.
  2008, \pasp, 120, 777

\bibitem[{{Vlahakis}(2006)}]{Vlahakis06}
{Vlahakis}, N. 2006, in Astronomical Society of the Pacific Conference Series,
  Vol. 350, Blazar Variability Workshop II: Entering the GLAST Era, ed. H.~R.
  {Miller}, K.~{Marshall}, J.~R. {Webb}, \& M.~F. {Aller}, 169

\bibitem[{{Walker} {et~al.}(2000){Walker}, {Dhawan}, {Romney}, {Kellermann}, \&
  {Vermeulen}}]{Walker_2000}
{Walker}, R.~C., {Dhawan}, V., {Romney}, J.~D., {Kellermann}, K.~I., \&
  {Vermeulen}, R.~C. 2000, \apj, 530, 233

\bibitem[{{Zamaninasab} {et~al.}(2013){Zamaninasab}, {Savolainen},
  {Clausen-Brown}, {Hovatta}, {Lister}, {Krichbaum}, {Kovalev}, \&
  {Pushkarev}}]{Zamanina13}
{Zamaninasab}, M., {Savolainen}, T., {Clausen-Brown}, E., {et~al.} 2013,
  \mnras, 436, 3341

\end{thebibliography}

\centering

\Online
\section{Online Material}

\begin{table*}
\caption{Model-fit parameters of 8\,GHz total intensity VLBI images.}   

\label{Table_2}      

 \resizebox*{0.9\textwidth}{!}{

\centering                          
\begin{tabular}{l c c c c c c c}    

 Date  &  Comp.  &  Flux  &  Rel. alpha &   Rel. delta    & Size                    &  Degree of         &  EVPA   \\ 
          &  		 &  	[Jy]	 &   [mas]    &  	[mas]          &  component [mas]   &   polarization [\%] &   [deg]    \\ 
         
\hline   

   2006-May-19  &   Q0   & 3.49   $\pm $  0.52  &  -0.11  $\pm $  0.13  &  0.04  $\pm $   0.10 &  0.34  $\pm $ 0.03  &  0.5  $\pm $  0.1  &  87.6  $\pm $ 5.3 \\
 		      &	   Q1    & 3.07  $\pm $  0.46  &  0.14   $\pm $  0.13  &  -0.05 $\pm $  0.10  &  0.46  $\pm $ 0.04  &  0.6  $\pm $ 0.1 &   87.3  $\pm $  5.4  \\
		     &	   Q5   & 0.12   $\pm $  0.02  &  2.32   $\pm $  0.66  &  2.42  $\pm $  0.50  &  1.93  $\pm $ 0.53  & 15.3 $\pm $  4.2  & 105.7 $\pm $ 7.8 \\

\hline 
  2006-November-23    	&  Q0  & 3.03   $\pm $  0.45   & -0.11 $\pm $  0.13  &  0.04   $\pm $  0.10  &  0.28 $\pm $  0.02 & 0.7   $\pm $  0.1  &  49.0  $\pm $  5.2 \\ 
 			  & Q1  & 2.98  $\pm $  0.44   & 0.15  $\pm $  0.13  &-0.04    $\pm $  0.10 & 0.40  $\pm $   0.04 &  0.7  $\pm $  0.1  &  55.4  $\pm $  5.8 \\
		       &    Q5  & 0.09  $\pm $  0.01 & 2.42  $\pm $  0.66   & 2.47    $\pm $  0.50  & 1.48 $\pm $   0.52  &  & \\
\hline    
 2008-January-12    &     Q0  & 4.07  $\pm $ 0.61 & -0.06    $\pm $   0.13   &  0.03 $\pm $ 0.10  & 0.36  $\pm $ 0.03 & 0.6     $\pm $ 0.1  &  35.2  $\pm $   5.3 \\
			& Q1   & 4.01   $\pm $ 0.60 &  0.10    $\pm $  0.13    &-0.02  $\pm $ 0.10  & 0.51  $\pm $ 0.05 &  0.5    $\pm $  0.1   & 40.2  $\pm $  6.5 \\
		       &   Q5 & 0.09    $\pm $ 0.01 &  2.20  $\pm $  0.66    & 2.05   $\pm $ 0.50  & 2.03  $\pm $ 0.54 & 12.2  $\pm $  3.3  & 119.3  $\pm $  6.7 \\

\hline 
 2009-January-17 & Q0  & 7.53 $\pm $  1.13 & -0.04     $\pm $  0.13 &  0.03   $\pm $ 0.10  & 0.39 $\pm $ 0.03 & 0.6       $\pm $ 0.1  &  68.9   $\pm $  6.6 \\
		  & Q1  & 4.11  $\pm $  0.61  & 0.12      $\pm $ 0.13  & -0.08 $\pm $ 0.10  & 0.59  $\pm $ 0.05 &  0.6      $\pm $  0.1  & 76.7  $\pm $  8.1 \\
		  &   Q5 & 0.08 $\pm $ 0.01 &  2.43   $\pm $ 0.66  &  2.5      $\pm $   0.50  & 1.95  $\pm $ 0.53 & 13.4  $\pm $ 2.0  & -60.4  $\pm $  5.7 \\ 

\end{tabular}
}
\end{table*}

\begin{table*}
\caption{Model-fit parameters of 15\,GHz total intensity VLBI images.}   

 \resizebox*{0.9\textwidth}{!}{
\label{Table_3}      
\centering                          
\begin{tabular}{l c c c c c c c}    

 Date  &  Comp.  &  Flux    &  Rel. alpha   & Rel. delta    & Size                         &  Degree of          & EVPA   \\ 
          &  		 &  	[Jy]	 &   [mas]    &	[mas]           &  component [mas]   &   polarization [\%]   &    [deg]        \\ 
         
\hline   
                    
                     
 2006-November-23 & Q0 &  2.26 $\pm $  0.34  & -0.07 $\pm $  0.07   &  0.16 $\pm $ 0.05 &  0.15  $\pm $ 0.01 &  1.1 $\pm $  0.1  &  75.3  $\pm $  5.0 \\
 		 &   Q1 & 1.77 $\pm $  0.26   &  0.16 $\pm $  0.07   &  0.01 $\pm $ 0.05 &  0.36  $\pm $ 0.30 & 1.0  $\pm $ 0.1   & 73.1  $\pm $  5.0 \\
		  &Q2 &   3.36 $\pm $  0.50   &  0.00 $\pm $  0.07 & -0.08 $\pm $ 0.05 &  0.11  $\pm $ 0.01  &  1.0  $\pm $ 0.1  &  73.0  $\pm $  5.0\\
		  & Q5 &  0.04 $\pm $  0.01 &  2.29 $\pm $  0.35   &  2.39 $\pm $ 0.27 &  1.42  $\pm $ 0.30  &   &\\

\hline 
 2007-May-10 & Q0  & 2.95  $\pm $ 0.44    &-0.05   $\pm $  0.07  &  0.11   $\pm $ 0.05  & 0.15   $\pm $ 0.01 & 0.4  $\pm $ 0.1  &   3.1   $\pm $  5.8 \\
 		 & Q1  & 1.90    $\pm $ 0.29    & 0.15   $\pm $  0.14  & 0.00 $\pm $  0.11     & 0.37  $\pm $ 0.14 & 0.5 $\pm $  0.8  &  14.5  $\pm $ 16.0 \\
 		 & Q2  & 3.30    $\pm $ 0.49    & 0.00 $\pm $ 0.07     & -0.08   $\pm $  0.05  & 0.12  $\pm $ 0.01 & 0.3 $\pm $  0.1  &   8.9   $\pm $  5.0 \\
  		& Q5  & 0.05     $\pm $ 0.01  & 2.35   $\pm $ 0.35     &  2.53   $\pm $ 0.27  & 1.61   $\pm $ 0.31 &  &  \\
 		\hline                       

   2008-January-12  &  Q0     & 7.11  $\pm $ 1.06   &   -0.02 $\pm $  0.07 &  0.05 $\pm $  0.05  & 0.18  $\pm $ 0.01 & 0.3  $\pm $ 0.2  &  22.1  $\pm $  14.5 \\
   			&  Q1  & 1.34  $\pm $ 0.20    &    0.18 $\pm $  0.14 & -0.04 $\pm $  0.11  & 0.40 $\pm $  0.14 & 0.4    $\pm $ 0.3  &  32.8  $\pm $  15.1 \\
 		     &  Q2  & 2.26  $\pm $ 0.34        &    0.02 $\pm $  0.07 &  -0.15 $\pm $  0.05  & 0.17  $\pm $  0.01 & 0.2 $\pm $  0.2  &  77.5  $\pm $ 16.7 \\
       		      &  Q5  & 0.07  $\pm $ 0.01       &    2.19 $\pm $  0.35 &    2.28  $\pm $ 0.27  &  2.48 $\pm $  0.37 &   &\\
\hline                       
2008-July-05  	 & Q0 & 7.61  $\pm $ 1.14 &  -0.02 $\pm $ 0.07  &  0.06  $\pm $ 0.05  & 0.19  $\pm $ 0.01 &  1.1 $\pm $  0.4  &  31.0  $\pm $  5.3 \\
  			& Q1 & 1.40  $\pm $ 0.21  &  0.17  $\pm $ 0.14  & -0.06  $\pm $ 0.11  & 0.38  $\pm $ 0.14 &  0.9 $\pm $  0.4  &  41.4  $\pm $  7.5 \\
   			 & Q2 & 2.62 $\pm $ 0.39  &  0.01  $\pm $ 0.07  & -0.16  $\pm $  0.05 &  0.16 $\pm $  0.01 &  0.6 $\pm $  0.1  &  35.2  $\pm $  5.0 \\
			&Q5 & 0.05  $\pm $ 0.01 &  2.16 $\pm $  0.35  &    2.21  $\pm $ 0.27  & 1.15  $\pm $ 0.29 &  &\\
   	        	 
\hline                       

 2009-January-17 & Q0 & 10.03 $\pm $ 1.50   &  -0.02 $\pm $  0.07 &  0.06 $\pm $  0.05 &  0.21  $\pm $ 0.02 & 2.4   $\pm $ 0.4  &   8.4  $\pm $  5.3 \\
    		 & Q1 &  0.99  $\pm $   0.15   &  0.22   $\pm $ 0.14 & -0.11 $\pm $  0.11  &  0.41  $\pm $ 0.14 & 1.3  $\pm $ 0.5  &  15.5 $\pm $  11.8 \\
  		  & Q2 &  3.52  $\pm $  0.52   &  0.03  $\pm $ 0.07  &-0.20  $\pm $ 0.05  &  0.22  $\pm $ 0.02 & 1.2   $\pm $ 0.2  &   4.3  $\pm $  6.5  \\
   		 & Q5 &  0.05  $\pm $   0.01 &  2.38  $\pm $ 0.35  & 2.22  $\pm $ 0.27  &  1.71  $\pm $ 0.32 &  & \\

\end{tabular}
}
\end{table*}

\begin{table*}
\caption{Model-fit parameters of 22\,GHz total intensity VLBI images.}   

 \resizebox*{0.9\textwidth}{!}{
\label{Table_4}      
\centering                          
\begin{tabular}{l c c c c c c c c}    
 Date  &  Comp.  &  Flux  & Rel. alpha   & Rel. delta    & Size                         &  Degree of         & EVPA    \\ 
          &  		 &  	[Jy]	  & [mas]       & 	[mas]          &  component [mas]   &   polarization [\%] &      [deg]    \\ 
         
\hline


2006-May-19  &			Q0&  2.32   $\pm $ 0.34    & -0.08  $\pm $ 0.03 &  0.17  $\pm $ 0.02  & 0.16 $\pm $  0.01 & 2.3  $\pm $  0.5  & -67.1  $\pm $  5.6 \\
 		 &			Q1&  0.72   $\pm $ 0.11    &   0.24  $\pm $ 0.06 &  0.05  $\pm $ 0.05 &  0.29 $\pm $  0.07 &  1.5 $\pm $  0.4 &   38.7  $\pm $ 18.0 \\
  				     &  Q2 &  3.27 $\pm $  0.49 &     0.00  $\pm $ 0.03 & -0.09  $\pm $ 0.02  & 0.10 $\pm $  0.01  & 0.4 $\pm $  0.1  & -14.5 $\pm $  11.9 \\
\hline   
2006-November-23  &  Q0  & 2.27  $\pm $  0.34  & -0.06  $\pm $   0.06  & 0.16   $\pm $ 0.05  & 0.18  $\pm $ 0.06 &  0.8  $\pm $  0.2  &  88.2  $\pm $  11.0 \\
  &			Q2 &  4.62 $\pm $   0.69 &  0.01  $\pm $  0.03 & -0.05  $\pm $  0.02 &  0.15 $\pm $  0.01  & 0.6 $\pm $ 0.2 &  -49.9   $\pm $  11.9 \\

\hline   
2007-May-10 & Q0 &  4.52   $\pm $  0.68  & -0.03 $\pm $  0.03 &  0.06  $\pm $  0.02 &  0.15 $\pm $  0.01  &  2.4 $\pm $  0.5  &  -3.6  $\pm $  7.0 \\
		&    Q2  & 3.34  $\pm $ 0.51  & 0.02    $\pm $ 0.06 & -0.10  $\pm $ 0.05 &  0.17  $\pm $ 0.06  &  1.1  $\pm $ 0.3  & -13.1  $\pm $  11.9 \\
\hline 

\hline 
2008-January-12  &  Q0   & 7.28   $\pm $ 1.09  &-0.01  $\pm $ 0.03 &  0.03  $\pm $ 0.02  & 0.16  $\pm $ 0.01 &  2.2 $\pm $  0.7 &  -12.5  $\pm $ 11.3 \\
  &			Q2  & 1.71   $\pm $ 0.26 &  0.05  $\pm $ 0.06 & -0.20  $\pm $ 0.05  & 0.16 $\pm $ 0.06 &  1.6 $\pm $  0.7 &  -32.4  $\pm $  8.1\\

\hline 

2008-July-05  &  Q0   & 7.63   $\pm $ 1.16   &-0.02   $\pm $ 0.06   & 0.05   $\pm $ 0.05   & 0.19   $\pm $ 0.06 & 2.1  $\pm $ 0.8  &   4.1    $\pm $  9.1 \\
  &			Q2  &  2.71  $\pm $  0.41  &  0.02  $\pm $  0.06  & -0.17  $\pm $  0.05  & 0.20   $\pm $ 0.06 & 1.5  $\pm $  0.5 &  144.1  $\pm $  6.1 \\
 
 \hline  
2009-January-17  &  Q0  & 5.51   $\pm $ 0.82    & -0.02  $\pm $  0.03 &   0.07 $\pm $  0.02 &   0.16 $\pm $   0.01 & 5.5 $\pm $  1.2 &   15.4 $\pm $   7.2 \\
&			Q2 &  0.54  $\pm $ 0.68  & 0.08   $\pm $ 0.03  &  -0.41   $\pm $ 0.02  &  0.10 $\pm $  0.01  & 3.4 $\pm $  0.5 &  154.8 $\pm $   5.1 \\
  &			Q3 &  4.50  $\pm $ 0.81    & 0.02   $\pm $ 0.06  &-0.11   $\pm $ 0.05  &  0.23  $\pm $ 0.07 & 2.1   $\pm $  0.9   & 19.3 $\pm $  13.3 \\

\end{tabular}
}
\end{table*}

\begin{table*}
\caption{Model-fit parameters of 43\,GHz total intensity VLBI images.}   

 \resizebox*{0.9\textwidth}{!}{
\label{Table_5}      
\centering                          
\begin{tabular}{l c c c c c c c }    
 Date  &  Comp.  &  Flux  & Rel. alpha  & Rel. delta    & Size                        &    Degree of         &  EVPA   \\ 
          &  		 &  	[Jy] & [mas]         & 	[mas]         &  component [mas]   &   polarization [\%] &     [deg]    \\ 
         
\hline


2006-May-19     &	Q0  & 1.24  $\pm $  0.19          &  -0.09 $\pm $  0.03 &  0.26     $\pm $   0.02 &    0.16  $\pm $    0.03 & 5.5  $\pm $ 1.6  &  91.3  $\pm $  11.9 \\
		      &		Q1  & 0.43  $\pm $  0.06 &            0.24 $\pm $   0.03 &  0.12     $\pm $   0.02 &   0.28  $\pm $   0.04  &                           & \\
			 &    Q2  & 1.58   $\pm $ 0.23  &           0.02   $\pm $    0.01  & -0.07  $\pm $ 0.01  &   0.05 $\pm $   0.01    & 1.6 $\pm $  0.3 &  138.5  $\pm $  10.0 \\
			&	Qn  & 1.55  $\pm $  0.23 &              -0.01 $\pm $   0.01 &  0.02     $\pm $   0.01 &   0.07 $\pm $   0.01 & 2.5  $\pm $ 0.6 &  128.4  $\pm $  10.1 \\

\hline  
2006-November-23   &	  Q0  & 1.16  $\pm $ 0.17 & -0.06  $\pm $ 0.03 &  0.20   $\pm $ 0.02   & 0.15  $\pm $ 0.03  & & \\
&			Q1  & 0.38   $\pm $ 0.05 &  0.26   $\pm $ 0.03 &  0.05   $\pm $ 0.02  & 0.25    $\pm $ 0.04 & & \\
&			Q2  & 1.14   $\pm $ 0.17 &  0.05   $\pm $ 0.01  &  -0.18 $\pm $  0.01 &  0.05 $\pm $ 0.01   & 3.5  $\pm $ 0.5  & -44.5 $\pm $   10.4 \\
&   			Qn  & 3.57   $\pm $ 0.53 &  0.00 $\pm $ 0.01  & -0.01  $\pm $  0.01 &  0.07 $\pm $ 0.01     & 2.9  $\pm $ 0.6 &  -42.8  $\pm $  11.6 \\

\hline  
2007-May-10    &		Q0 &  0.68    $\pm $  0.10 & -0.05    $\pm $ 0.03 &  0.21 $\pm $ 0.02   & 0.14  $\pm $ 0.03     & 6.2  $\pm $ 1.4  & -13.5  $\pm $  10.3 \\
	&			Q1 &  0.25    $\pm $  0.03 &  0.29    $\pm $ 0.03 & -0.01 $\pm $ 0.02   & 0.20  $\pm $ 0.04    & &  \\
	&			Q2 &  1.07    $\pm $  0.16 &  0.06    $\pm $ 0.01 & -0.23 $\pm $ 0.01   &  0.06 $\pm $  0.01    & & \\
				&  Qn & 4.81 $\pm $ 0.73 &  0.00  $\pm $ 0.03 &     0.00 $\pm $ 0.02    & 0.10 $\pm $ 0.03 & 4.6 $\pm $  1.1  & 147.8  $\pm $  11.6 \\

\hline  
2008-January-12    &		Q0 &  2.98 $\pm $  0.45 &  0.01 $\pm $ 0.03 &  0.05  $\pm $ 0.02 & 0.10  $\pm $ 0.03 & 7.6 $\pm $  2.5  &  -2.1  $\pm $  13.7 \\
			&	Q1 &  0.18 $\pm $  0.03 &  0.31 $\pm $ 0.08 & -0.01  $\pm $ 0.06  & 0.25 $\pm $ 0.06 & & \\
			&	Q2 &  1.17 $\pm $  0.17 &  0.06 $\pm $ 0.03 & -0.29  $\pm $ 0.02 & 0.08  $\pm $ 0.03 & & \\
			&  Qn  &  4.00   $\pm $ 0.61  & -0.01  $\pm $ 0.03 &  -0.03 $\pm $ 0.02  & 0.12  $\pm $  0.03 & 2.8 $\pm $  0.9  & -37.7 $\pm $  26.0 \\

\hline  

2008-July-05    &	Q0 &  4.23     $\pm $   0.64  & 0.00  $\pm $  0.03  &  0.06  $\pm $  0.02 &  0.15  $\pm $   0.03 &  4.2  $\pm $ 1.9  &  -11.3  $\pm $ 17.8\\
			& Q1 &  0.31  $\pm $  0.04  &  0.22    $\pm $   0.03  & -0.08  $\pm $ 0.02 &  0.24  $\pm $  0.04  &  8.7  $\pm $ 3.1  &  51.5   $\pm $ 11.0 \\
			& Q2 &  0.66  $\pm $  0.09 &   0.07   $\pm $  0.01  & -0.33    $\pm $  0.01 &   0.07 $\pm $   0.01 & &  \\
			&  Q3  & 2.40 $\pm $  0.36  & 0.00    $\pm $  0.03  & -0.10  $\pm $  0.02 &   0.15  $\pm $  0.03 & 3.0  $\pm $ 1.0  & 132.1  $\pm $  10.9\\
\hline 
2009-January-17  &  Q0   &4.80  $\pm $ 0.73    &-0.01  $\pm $ 0.03 &  0.06   $\pm $  0.02 &  0.15           $\pm $  0.03    & 9.0     $\pm $  1.9 &  -7.2 $\pm $  13.4 \\
&			Q1  & 0.10  $\pm $ 0.01  & 0.35   $\pm $ 0.03  &-0.13 $\pm $ 0.02  & $\leq $ 0.03                          & &\\
&			Q2  & 0.89  $\pm $ 0.13  & 0.08   $\pm $ 0.03 & -0.46 $\pm $ 0.02  &  0.10             $\pm $ 0.03 & & \\
&			Q3  & 3.44  $\pm $ 0.52  & 0.02   $\pm $ 0.03 & -0.12 $\pm $ 0.02  &  0.19              $\pm $ 0.03 &  3.2         $\pm $ 1.3 &  5.8 $\pm $   38.1 \\

\end{tabular}
}
\end{table*}

\begin{table*}
\caption{Model-fit parameters of 86\,GHz total intensity VLBI images.}   

 \resizebox*{0.9\textwidth}{!}{
\label{Table_6}      
\centering                          
\begin{tabular}{l c c c c c c c }    
 Date  &  Comp.  &  Flux  & Rel. alpha   & Rel. delta   & Size                         &  Degree of           & EVPA   \\ 
          &  		 &  	[Jy]  & [mas]         & 	[mas]        &  component [mas]   &   polarization [\%]    &    [deg]       \\ 
         
\hline


2006-May-07   &		Q0&  0.14   $\pm $  0.02 & -0.11  $\pm $  0.02 &  0.34   $\pm $ 0.01 & 0.10 $\pm $  0.02 \\
 			   &   Q2 & 0.65  $\pm $ 0.10  & 0.00  $\pm $ 0.02   &  0.00  $\pm $0.01  & 0.05 $\pm $  0.02 \\
	&			Qn&  0.21  $\pm $  0.03 & -0.03 $\pm $  0.02  &  0.11   $\pm $ 0.01 & 0.06 $\pm $  0.02 \\

\hline  
2008-May-09   &	Q0 &  1.20    $\pm $ 0.18 &    0.00 $\pm $  0.01 &  0.00 $\pm $  0.01 & $\leq $ 0.02  \\
 		&      Q3   & 0.38   $\pm $0.06 &  0.06     $\pm $ 0.06   &-0.20  $\pm $ 0.02  &            0.15  $\pm $ 0.03 \\
\hline  
2009-May-08  &   Q0 & 2.30 $\pm $ 0.35 &  0.00  $\pm $ 0.02 &      0.01 $\pm $  0.01  &   0.09  $\pm $ 0.02  &   7.4  $\pm $  3.1   & 0.3  $\pm $  22.8 \\
&			Q2 & 0.61 $\pm $  0.09&   0.05    $\pm $  0.02 &  -0.54 $\pm $   0.01 &  0.03  $\pm $  0.02  &  8.3  $\pm $  1.3  &  13.7  $\pm $ 10.0 \\
&			Q3 & 0.65 $\pm $ 0.10 &  0.02   $\pm $ 0.02 &    -0.17 $\pm $  0.01  &    0.13   $\pm $ 0.02 \\
\hline  
2010-May-07  &   Q0  & 2.60  $\pm $ 0.39 &  0.00 $\pm $ 0.02  & -0.03 $\pm $  0.01 & 0.12 $\pm $ 0.02  & 9.2    $\pm $  4.4  &  -5.6  $\pm $  12.2 \\
&			Q2  & 1.55  $\pm $ 0.23  & 0.09 $\pm $ 0.02  & -0.61   $\pm $ 0.01  & 0.10 $\pm $ 0.02 &  7.9  $\pm $  1.8  &  15.6  $\pm $ 11.2 \\
&			Q3  & 0.40  $\pm $ 0.07  & 0.01 $\pm $ 0.02  & -0.29   $\pm $  0.01  & 0.07 $\pm $  0.02 \\
\hline  
 



\end{tabular}
}
\end{table*}


\end{document}